\documentclass[reprint,aps,prapplied,amssymb,twocolumn,showpacs,floatfix,superscriptaddress,nobalancelastpage,footinbib]{revtex4-1}

\usepackage{graphicx}
\usepackage{multirow}
\usepackage{amssymb}
\usepackage{amsmath}
\usepackage{color}
\usepackage{url}
\usepackage{epstopdf}
\usepackage{mathrsfs}

\usepackage[breaklinks, urlcolor=blue, hyperindex, colorlinks, bookmarks=true]{hyperref}

\usepackage{microtype}

\begin{document}

\title{Mode Structure in Superconducting Metamaterial Transmission Line Resonators}


\author{H. Wang}
\affiliation{Department of Physics, Syracuse University, Syracuse, NY 13244-1130}

\author{A.P. Zhuravel}
\affiliation{B. Verkin Institute for Low Temperature Physics \rm{\&} \textit{Engineering of
National Academy of Science of Ukraine,
UA-61103 Kharkiv, Ukraine}}

\author{S. Indrajeet}
\affiliation{Department of Physics, Syracuse University, Syracuse, NY 13244-1130}

\author{B.G. Taketani }
\affiliation{Departamento de F\'{i}sica, Universidade Federal de Santa Catarina, 88040-900 Florian\'{o}polis, SC, Brazil}
\affiliation{Theoretical Physics, Saarland University, Campus, 66123 Saarbr\"{u}cken, Germany}

\author{M.D. Hutchings}
\altaffiliation{Present address: SeeQC, Inc., Suite 141, 175 Clearbrook Road, Elmsford, NY 10523, USA}
\affiliation{Department of Physics, Syracuse University, Syracuse, NY 13244-1130}

\author{Y. Hao}
\altaffiliation{Present address: SiTime Corporation
5451 Patrick Henry Drive,
Santa Clara, CA 95054
USA l}
\affiliation{Department of Physics, Syracuse University, Syracuse, NY 13244-1130}

\author{F. Rouxinol}
\altaffiliation{Present address: ``Gleb Wataghin" Institute of Physics,
University of Campinas – UNICAMP
13083-859, Campinas, SP, Brazil}
\affiliation{Department of Physics, Syracuse University, Syracuse, NY 13244-1130}

\author{F.K. Wilhelm}
\affiliation{Theoretical Physics, Saarland University, Campus, 66123 Saarbr\"{u}cken, Germany}

\author{M.D. LaHaye}
\affiliation{Department of Physics, Syracuse University, Syracuse, NY 13244-1130}

\author{A.V. Ustinov}
\affiliation{Physikalisches Institut, Karlsruhe Institute of Technology, 76131 Karlsruhe, Germany }

\affiliation{Russian Quantum Center, National University of Science and Technology MISIS, 119049, Moscow, Russia}

\author{B.L.T. Plourde}
\email[]{bplourde@syr.edu}
\affiliation{Department of Physics, Syracuse University, Syracuse, NY 13244-1130}

\date{\today}


\begin{abstract} 
Superconducting metamaterials are a promising resource for quantum information science. In the context of circuit QED, they provide a means to engineer on-chip, novel dispersion relations and a band structure that could ultimately be utilized for generating complex entangled states of quantum circuitry, for quantum reservoir engineering, and as an element for quantum simulation architectures.  Here we report on the development and measurement at millikelvin temperatures of a particular type of circuit metamaterial resonator composed of planar superconducting lumped-element reactances in the form of a discrete left-handed transmission line (LHTL) that is compatible with circuit QED architectures.  We discuss the details of the design, fabrication, and circuit properties of this system. As well, we provide an extensive characterization of the dense mode spectrum in these metamaterial resonators, which we conducted using both microwave transmission measurements and laser scanning microscopy (LSM). Results are observed to be in good quantitative agreement with numerical simulations and also an analytical model based upon current-voltage relationships for a discrete transmission line. In particular, we demonstrate that the metamaterial mode frequencies, spatial profiles of current and charge densities, and damping due to external loading can be readily modeled and understood, making this system a promising tool for future use in quantum circuit applications and for studies of complex quantum systems.
\end{abstract}

\maketitle

\section{Introduction}
\label{sec:Introduction}


Engineered quantum systems present a unique opportunity to study many-body phenomena in a controlled way through analog quantum simulation \cite{Cirac12}.
An intriguing direction in this area involves the strong coupling of atoms -- artificial or natural -- to a quasicontinuum of modes, as is the case with quantum impurity models or the investigation of scattering in strongly interacting systems \cite{Mostame2016}. 
%

In the microwave regime, the field of circuit QED (cQED) involves one or more artificial nonlinear few-level systems (qubits) that are coupled to linear resonant modes, often formed from standing waves on a finite-length microwave transmission line~\cite{Blais2004} or a three-dimensional waveguide cavity~\cite{Paik11}. This platform permits the attractive feature of reaching strong coupling between the qubit and individual photonic modes in the resonator due to both the large transition dipole of the qubit and the small mode volume of the resonator, which is now utilized regularly for experiments in quantum-information processing and quantum optics in the microwave regime~\cite{blais2007quantum,Majer07,gu2017microwave,Braum2017}. 

Recent efforts have been undertaken to achieve multimode strong coupling in cQED systems, wherein a superconducting qubit is able to couple strongly to multiple photonic modes. Efforts to engineer a spectrally dense mode structure include the use of electrically long (${\rm approximately} 1\,{\rm m}$) superconducting transmission lines \cite{sundaresan2015beyond}, arrays of coupled transmission-line resonators or cQED systems \cite{underwood2012low, fitzpatrick2017observation,Mckay2015, naik2017random,liu2017quantum,Lepp2018}, and frequency combs implemented through a pumped superconducting nonlinearity \cite{Erickson2014}.  

An alternative approach, which we explore  in a cQED architecture here, utilizes a periodic array of superconducting lumped-element reactances to engineer a one-dimensional metamaterial in the microwave range \cite{Egger13,messinger2018left}.   This  metamaterial is characterized by a photonic bandgap at low frequencies, with a band characterized by a left-handed dispersion relation \cite{Eleftheriades2002,Caloz2004tl,Sanada2003,caloz2004novel} and a dense set of modes at frequencies just above the bandgap, commensurate with superconducting qubit transition energies.  A variety of superconducting metamaterials have been studied previously \cite{wang2006high,Jung2014multi,zagoskin2016quantum,hutter2011josephson,jung2014progress}, including some systems with applications in microwave quantum optics \cite{martinez2018probing,mirhosseini2018superconducting} and quantum-limited amplification of microwave signals \cite{castellanos2008amplification,macklin2015near,zhang2017josephson}. In our present work, the dense mode spectrum we implement is appealing for future applications for architectures of quantum simulators and entanglement generation \cite{Egger13}, whereby the qubits could be tuned between the low-frequency bandgap and the left-handed region of the spectrum just above the bandgap. However, in contrast to earlier work on multimode cQED systems, which involved physically long cavities or coupled distributed transmission lines, our use of lumped-element components provides a more compact design that can be readily interfaced with superconducting qubits, and thus is inherently scalable for applications in quantum information and communication.

Here we present a characterization of the mode structure in superconducting metamaterial resonators in a cQED architecture that is compatible with superconducting qubit integration through low-temperature measurements of transmission spectra, imaging of microwave fields with a laser-scanning microscope  (LSM), as well as numerical simulations. Our results lay the foundations for future applications of this system for quantum simulations and the design of  superconducting resonator spectra.  In Sec.~\ref{sec:1D Metamaterial transmission lines} we discuss the general structure of 1D metamaterial transmission lines and resonators. We describe the fabrication and microwave characterization of our superconducting metamaterial resonators in Sec.~\ref{sec:Fab}. In Sec.~\ref{sec:LSM imaging of mode structure} we present our investigation of the mode structure through LSM imaging of the microwave fields, which we analyze in detail in Sec.~\ref{sec:analysis of LSM}. We present numerical simulations of our metamaterial design in Sec.~\ref{sec:Sonnet}. In Sec.~\ref{sec:loss} we present an analysis of losses in superconducting metamaterial resonators. Section \ref{sec:conclusions} contains our conclusions and possible future directions for experiments with superconducting metamaterial resonators.

\section{1D Metamaterial transmission lines}
\label{sec:1D Metamaterial transmission lines}



\begin{figure}[b]
\centering
\includegraphics[width=3.35in]{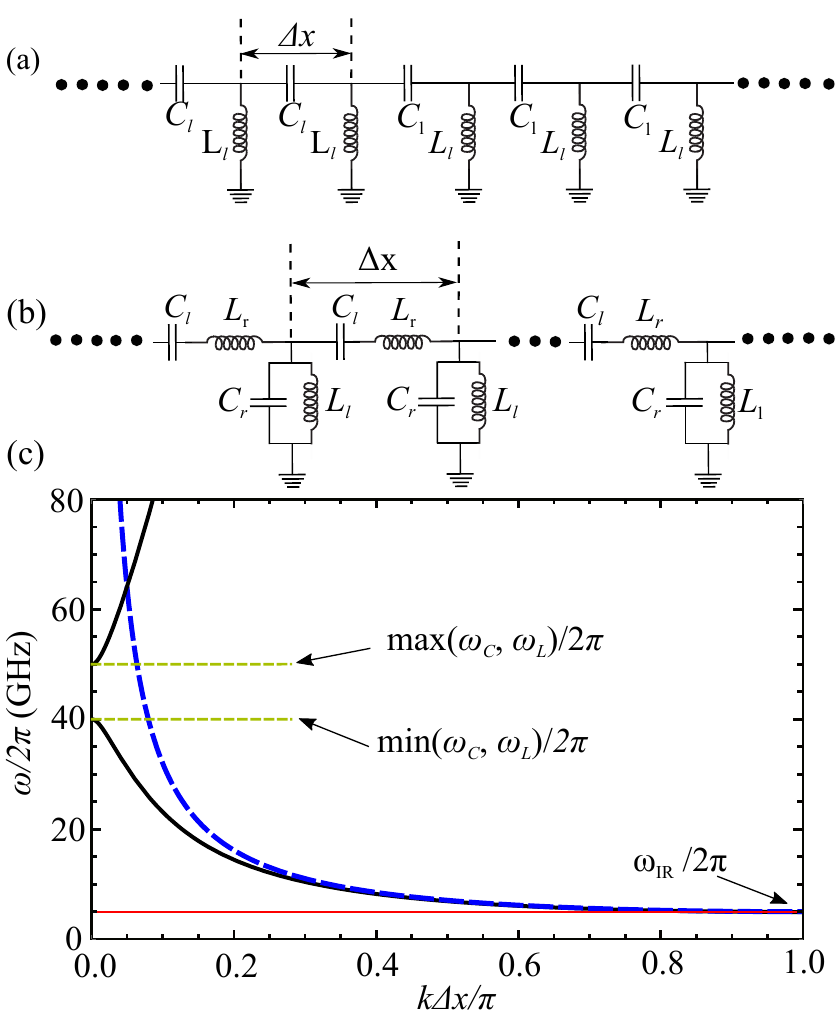}
  \caption{Metamaterial transmission line schematic: (a) ideal LHTL; (b) composite LHTL including stray reactances; (c) calculated dispersion relations $\omega(k)$ for ideal LHTL (dashed blue line) [Eq.~(\ref{eq:ideal-LHL-disp})] and composite LHTL (solid black line) [Eq.~(\ref{eq:Caloz-CRLH-disp})] using circuit parameters described in the text. 
\label{fig:meta-schem}}
\end{figure}

The basic properties of 1D left-handed transmission lines (LHTL) have been treated previously by several others \cite{Eleftheriades2002,Iyer09,caloz2004novel,Egger13,Ovchinnikova2013}, thus, we only review these briefly. 
The simplest circuit model for a LHTL consists of a chain of capacitors $C_l$ and inductors $L_l$ in a similar manner to the discrete implementation of a conventional right-handed transmission line that would be described by the telegrapher's equation \cite{Pozar-book}. However, the LHTL has the positions of the inductors and capacitors swapped with respect to the right-handed line [Fig.~\ref{fig:meta-schem}(a)]. 
This rearrangement has a profound effect on the transmission and dispersion properties of the transmission line.

The dispersion relation for such an LHTL is given by 
\begin{equation}
\omega(k) = \frac{1}{2 \sqrt{L_l C_l}}\frac{1}{\sin \left( k \Delta x/2 \right)},
\label{eq:ideal-LHL-disp}
\end{equation}
where $k$ is the wavevector in the LHTL and $\Delta x$ is the unit cell length \cite{Egger13}. This expression is plotted in Fig.~\ref{fig:meta-schem}(c) and the curve clearly exhibits a falling dispersive behavior, characteristic of a left-handed medium. At the largest wave numbers in the limit of $k \Delta x \rightarrow \pi$, or one wavelength for every two unit cells, corresponding to the lowest frequencies of the system, the band becomes flat approaching the edge of the Brillouin zone. The corresponding frequency, $\omega_{IR}=1/2\sqrt{L_l C_l}$, defines a high-pass infrared cutoff frequency below which the structure does not support propagating waves. As the wavenumber is decreased the frequency diverges, which is clearly unphysical as this implies that the group velocity diverges in the limit of $k \rightarrow 0$.

Such behavior would not occur in any practical transmission line due to the stray reactances in the constituent elements for an LHTL, as described in Refs.~\cite{caloz2005book, Lai2004Caloz, caloz2004novel}. Each capacitor $C_l$ will have a stray series inductance $L_r$ giving it a self-resonance frequency $\omega_C = 1/\sqrt{L_r C_l}$. Similarly, each inductor $L_l$ will have a stray shunt capacitance $C_r$ resulting in another self-resonance frequency $\omega_L = 1/\sqrt{L_l C_r}$. Figure~\ref{fig:meta-schem}(b) shows a schematic of a composite LHTL including these stray reactances. As first described in Ref.~\cite{caloz2005book}, 
a Bloch-Floquet analysis can be applied to such a composite LHTL to obtain the dispersion relation between wavenumber $k$ and frequency $\omega$. We include our own derivation in Appendix~\ref{appendix:Derivation of dispersion relation} and obtain the following expression~\cite{Lai2004Caloz, caloz2004novel}: 
\begin{equation}
\begin{split}
k(\omega)=&\frac{1}{\Delta x}\times\\
&\cos^{-1}\left[1-\frac{1}{2}\left(\omega L_r-\frac{1}{\omega C_l}\right)\left(\omega C_r-\frac{1}{\omega L_l}\right) \right].
\end{split}\label{eq:Caloz-CRLH-disp}
\end{equation}
%

Figure~\ref{fig:meta-schem}(c) includes a plot of $\omega(k)$ obtained by inverting Eq.~(\ref{eq:Caloz-CRLH-disp}). Similar to the corresponding curve from Eq.~(\ref{eq:ideal-LHL-disp}), the band becomes flat as $k \Delta x \rightarrow \pi$ at $\omega_{IR}/2\pi$, which is chosen to be $5\,{\rm GHz}$ for this plot. However, for small $k \Delta x$ the curve does not diverge and instead turns over and intersects $k \Delta x=0$ at the lower of the two stray self-resonance frequencies, $\min(\omega_C,\omega_L)$. Nonetheless, the curve still has a left-handed nature over the entire range of $k \Delta x$. At the upper frequency end of the plot, it is clear that there is a second band corresponding to another solution of Eq.~(\ref{eq:Caloz-CRLH-disp}). This upper band begins at the higher of the two self-resonance frequencies, $\max(\omega_C,\omega_L)$, and increases with $k$, characteristic of right-handed dispersion. Between $\omega_C$ and $\omega_L$ there is a gap where there are no solutions to Eq.~(\ref{eq:Caloz-CRLH-disp}) and thus no wave propagation along the transmission line. For this plot, estimates of likely self-resonance values in physically realizable structures are used: $\omega_C/2\pi=40\,{\rm GHz}$, $\omega_L/2\pi=50\,{\rm GHz}$. For the remainder of the discussion we  focus on the left-handed branch, as the  devices  we study put the high-frequency gap and the right-handed branch well beyond the frequency range that we are able to access in our measurements. However, we note that with future devices we may be able to enhance the stray reactances in order to lower the frequency of the gap and bring the right-handed branch into our measurement range.

\begin{figure}
\centering
\includegraphics[width=3.35in]{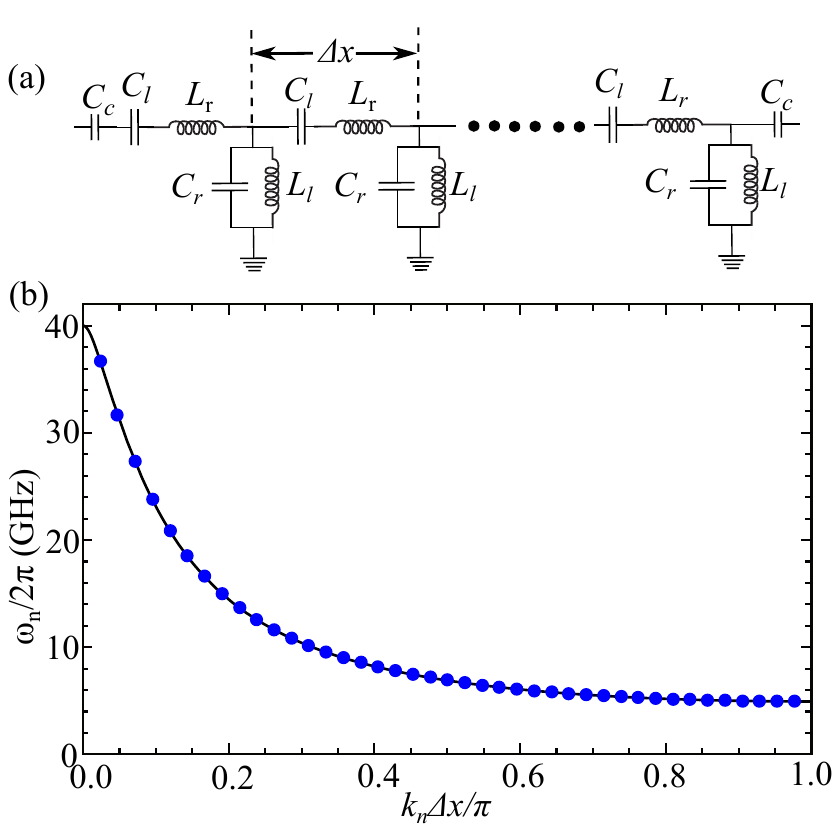}
  \caption{(a) Schematic of a composite LHTL resonator with stray reactances including coupling capacitances $C_c$ at each end. (b) Plot of mode frequencies vs. wave number computed for $N-1=41$ modes in a LHTL ($\omega_{IR}/2\pi$=5\,{\rm GHz}, $\omega_C/2\pi=40\,{\rm GHz}$, $\omega_L/2\pi=50\,{\rm GHz}$). The solid line is the disperson relation obtained from Eq.~(\ref{eq:Caloz-CRLH-disp}). 
\label{fig:CRLH-res}}
\end{figure}

In order to generate resonances with the transmission line, one can impose either open-ended or short-circuited boundary conditions at two positions along the length of the line. We choose to focus on the case with open-ended boundary conditions, using input and output coupling capacitors $C_c$ at either end. As with a conventional right-handed transmission line, the system will exhibit standing-wave resonances given by the standard relation $k_n l = n \pi$ for integer $n$, where $l$ is the length of the transmission line between the two ends. With a discrete LHTL including stray reactances with $N$ unit cells, the standing-wave condition becomes $k_n \Delta x = n\pi/N$ for $n=0:N-1$ \cite{Sanada2003}. Thus, if we disregard the zeroth-order mode $(n=0)$ \cite{Sanada2003}, there are$N-1$ modes for frequencies between $\omega_{IR}$ and $\min(\omega_C, \omega_L)$. Due to the shape of the dispersion relation given by Eq.~(\ref{eq:Caloz-CRLH-disp}), the lower frequency and higher $n$ modes areclosely spaced in frequency with the mode spacing increasing as one moves towards higher frequency and lower $n$. This behavior is illustrated in the plot of Fig.~\ref{fig:CRLH-res}(b), which shows $\omega_n(k_n)$ points for each of the $N-1$ left-handed modes computed for a circuit with the same parameters as that considered in Fig.~\ref{fig:meta-schem}.

\section{Fabrication and characterization of metamaterial resonators}
\label{sec:Fab}

In order to implement metamaterial resonators that are compatible with the integration of superconducting transmon qubits \cite{Koch2007} that can be tuned through resonance with the various modes in the dense region of the metamaterial spectrum just above $\omega_{IR}$, we aim for $\omega_{IR}/2\pi \sim 4-7\,{\rm GHz}$. If we target a characteristic impedance near $50\,\Omega$, this requires $C_l=250\,{\rm fF}$ and $L_l=0.625\,{\rm nH}$. These parameter values can be achieved with superconducting thin-film structures consisting of interdigitated capacitors and meander-line inductors with a single-layer fabrication process that can easily be extended to incorporate superconducting qubits in future devices. Based on numerical simulations of these structures with Sonnet~\cite{suitesversion}, we expect the stray reactances $L_r$ and $C_r$ to result in self-resonant frequencies $\omega_C/2\pi$ and $\omega_L/2\pi$ of $40\,{\rm GHz}$ or higher, well beyond the frequency range of our measurements. Thus, we expect our metamaterial resonators to exhibit a left-handed dispersion over the entire range of our measurements. While some of our initial devices have been fabricated from Al thin films on Si \cite{plourde2015superconducting}, we  fabricate the metamaterial resonators described in this work using $90$-nm-thick Nb  films on Si. The higher $T_c$ of the Nb (approximately $9\,{\rm K}$) makes the structures compatible with the typical operating temperature of our microwave imaging technique, which is described in Sec.~\ref{sec:LSM imaging of mode structure}. The features are defined using  photolithography with a standard liftoff process following the Nb sputter deposition.

\begin{figure}
\centering
\includegraphics[width=3.35in]{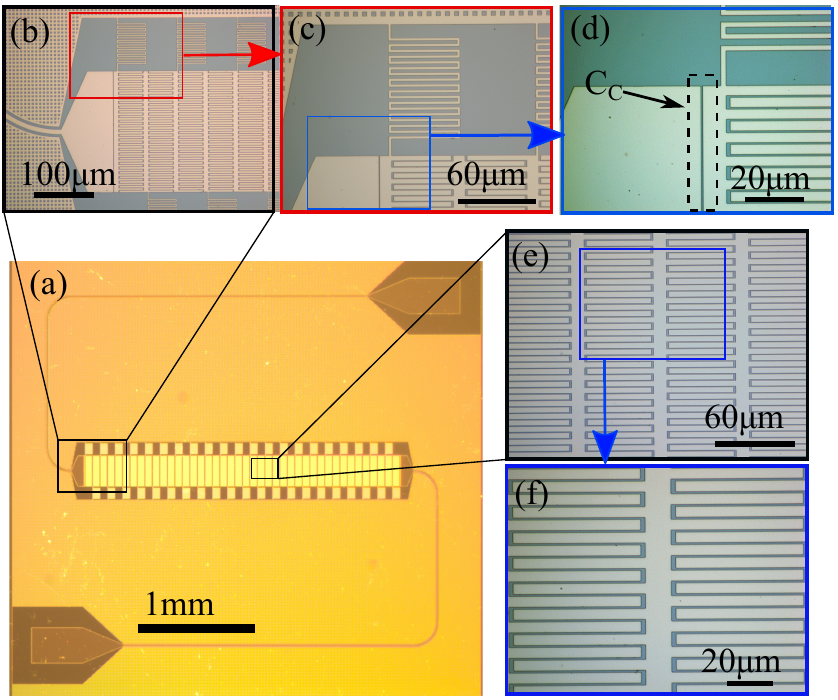}
  \caption{Optical micrographs of metamaterial resonator device: (a) zoomed-out image of entire chip, (b) input coupling capacitor $C_c$ and the first few unit cells, (c) meander-line inductor of the first unit cell, (d) detail of input coupling capacitor and connection between inductor and capacitor of the first unit cell, (e) interdigitated capacitors in several unit cells, (f) detail of interdigitated capacitor.
\label{fig:fab}}
\end{figure}

The meander-line inductors have a $2$-$\mu{\rm m}$ linewidth and 17 turns [Fig.~\ref{fig:fab}(c)]. The interdigitated capacitors have 26 pairs of 50-$\mu$m-long, 4-$\mu$m-wide fingers spaced by $1~\mu{\rm m}$ [Figs.~\ref{fig:fab}(e, f)]. The metamaterial transmission line runs across most of the width of the chip, which has an active area of $4 \times 4\,{\rm mm}^2$ [Fig.~\ref{fig:fab}(a)] when mounted in one of our standard microwave chip holders. Based on the length of the interdigitated capacitor fingers, this allows us to fit $N=42$ unit cells on a chip. The ends of the metamaterial resonator are defined with coupling-capacitor gaps at the first and last cells [Fig.~\ref{fig:fab}(d)] to conventional coplanar waveguide (CPW) traces carrying the output (input) signals to (from) launcher pads at the opposite corners of the chip. We estimate the $C_c$ values for these structures to be $30\,{\rm fF}$ based on HFSS Q3D simulations~\cite{HFSS}.

Because of the spatial extent of the meander-line inductors, we choose to stagger the placement of the inductors so that alternating inductors connect to the ground plane either upwards or downwards from the path of the transmission line [Fig.~\ref{fig:fab}(a)]. Besides allowing the inductors to fit in the available space, this configuration also avoids the asymmetry in the grounding that would occur if all of the inductors extended to ground on only one side of the transmission line. However, as we  show later, the trade-off of using the staggered inductor arrangement results in a distortion of the dispersion relation from that described by Eq.~(\ref{eq:Caloz-CRLH-disp}) for the lowest frequency modes.


For initial measurements of the mode spectrum, we cool the device on an adiabatic demagnetization refrigerator (ADR) with a base temperature of $65$~mK. The device is shielded from stray magnetic fields with a cryogenic $\mu$-metal can and microwave signals are delivered to the metamaterial through a coaxial line with 74~dB of cold attenuation; the output signal is amplified with a cryogenic HEMT plus a room-temperature low-noise microwave amplifier. We use a vector network analyzer to measure the microwave transmission $S_{21}(f)$. A separately measured baseline transmission curve with no sample present is used to normalize the $S_{21}(f)$ spectra.

\begin{figure}[htb]
\centering
\includegraphics[width=3.35in]{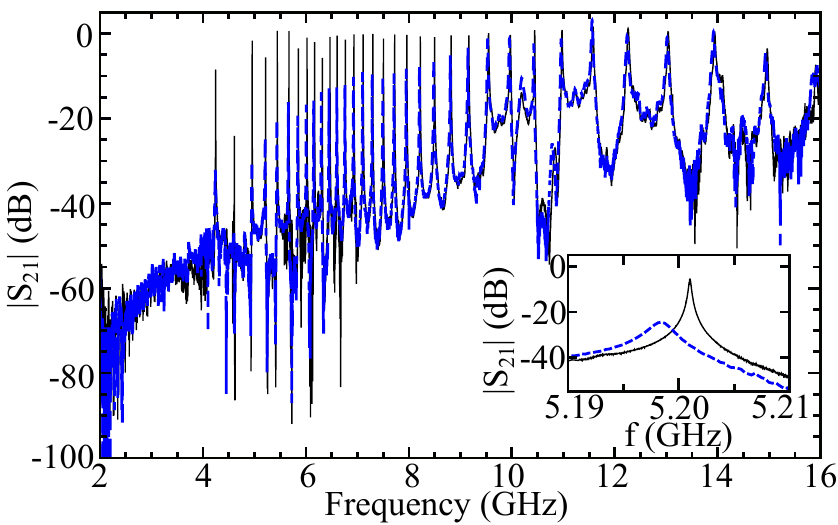}
  \caption{Measurements of the magnitude of the microwave transmission $|S_{21}(f)|$ on the ADR at two different temperatures: 65~mK (solid black line); 3~K (blue dashed line). The inset shows an enlarged plot in the vicinity of $n=38$ mode near 5.21~GHz.
\label{fig:T-dep}}
\end{figure}

Figure~\ref{fig:T-dep} contains a plot of our measured $S_{21}(f)$ spectra for one of our devices at two different temperatures: 65~mK and 3~K. Below 4.2~GHz, the transmission is below -40~dB, characteristic of the IR cutoff expected for a LHTL. For frequencies above this cutoff, we observe densely spaced narrow resonances that become increasingly far apart with broader linewidths at higher frequencies. The frequency placement of the resonances are discussed in Sec.~\ref{sec:analysis of LSM} and the coupling-limited linewidths are analyzed in Sec.~\ref{sec:loss}. When the sample temperature is raised to 3~K, approaching the operating temperature range for the LSM imaging, the resonances are still present, although shifted slightly lower in frequency due to the increased kinetic inductance, with reduced quality factors because of the presence of thermal quasiparticles.

\section{LSM imaging of mode structure}
\label{sec:LSM imaging of mode structure}

Low-temperature laser-scanning microscopy (LSM) has been used previously to image a wide variety of superconducting structures under rf excitation \cite{kurter2011,zhuravel2006}. The various modes of LSM operation have been described in detail in Refs.~\cite{zhuravel2006laser} and \cite{culbertson1998optical}.
The basic principle of operation of the LSM is based on the local deposition of laser energy onto a superconducting sample followed by the measurement of some global response of the sample, for example $S_{21}$, to the perturbation at that particular location. By raster scanning the laser probe over the surface of the sample while modulating the beam intensity, an image of the photoresponse $\mathscr{R}(x,y,f)$ 
can be produced using a lock-in technique by correlating the output signal with the location of the laser spot.


For the metamaterial imaging reported here, the chip is mounted inside a vacuum volume of a cryostat with optical access and the temperature of the chip is measured to be about $\,5$ K during the imaging. The laser probe is produced by a diode laser with wavelength $640\,{\rm nm}$ that is focused to a spot of about $12$-$\mu{\rm m}$ diameter with a power of approximately $10\,\mu{\rm W}$ at the sample, resulting in an increase of local temperature  at this location. The laser intensity is modulated at 100~kHz while the laser-induced modulation of the microwave transmission $|\Delta S_{21}(f)|$ is measured with cryogenic coaxial cables carrying microwave signals to and from the sample. The output signal is amplified by 60~dB, rectified with a crystal diode, then lock-in detected at the modulation frequency to generate the photoresponse signal.
The maximum frequency accessible on our LSM is $20\,{\rm GHz}$, similar to the measurement electronics for the experiments on our ADR.
\begin{figure}[htb]
\centering
\includegraphics[width=3.35in]{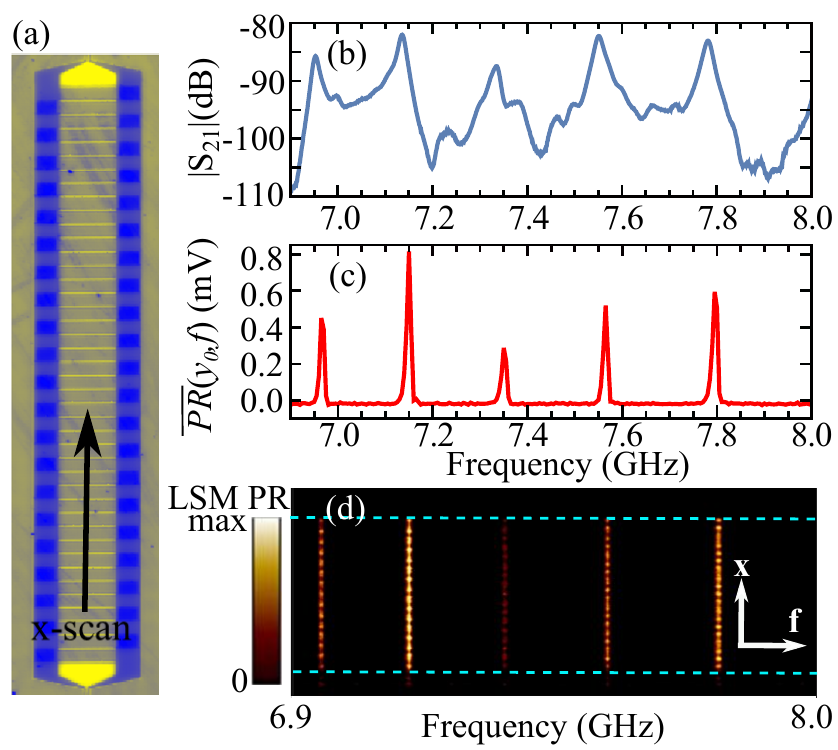}
  \caption{(a) LSM reflectivity image with the arrow indicating the direction of 1D line scans. (b) Microwave transmission (not normalized) $|S_{21}(f)|$ measured on LSM. (c) Average LSM photoresponse $\overline{\mathscr{R}}(y_0, f)$ along 1D line scans. (d) 1D line scans $\mathscr{R}(x,y_0,f)$ vs. frequency; dashed horizontal lines indicate location of input and output coupling capacitors.
\label{fig:LSMS21}}
\end{figure}


Figure~\ref{fig:LSMS21}(a) contains a reflectivity image of the metamaterial in the LSM, with an arrow indicating the orientation and location of 1D scans along the $x$ axis of the photoresponse that are measured in the LSM.
As the frequency is varied, the photoresponse signal $\overline{\mathscr{R}}(y_0, f)$ averaged along the 1D line scans exhibited sharp peaks [Fig.~\ref{fig:LSMS21}(c)], which line up  with the microwave transmission resonances in $|S_{21}(f)|$ that is measured simultaneously [Fig.~\ref{fig:LSMS21}(b)]. In Fig.~\ref{fig:LSMS21}(d), the density plot of $\mathscr{R}(x, y_0, f)$, where $y_0$ indicates the location of the line scan, again shows the sharp features at the resonant frequencies, but also exhibits fine structure along the scan direction. Thus, it is clear that the LSM imaging of the device can be used to investigate resonances in the metamaterial and their corresponding standing-wave patterns.
\begin{figure}[htb]
\centering
\includegraphics[width=3.35in]{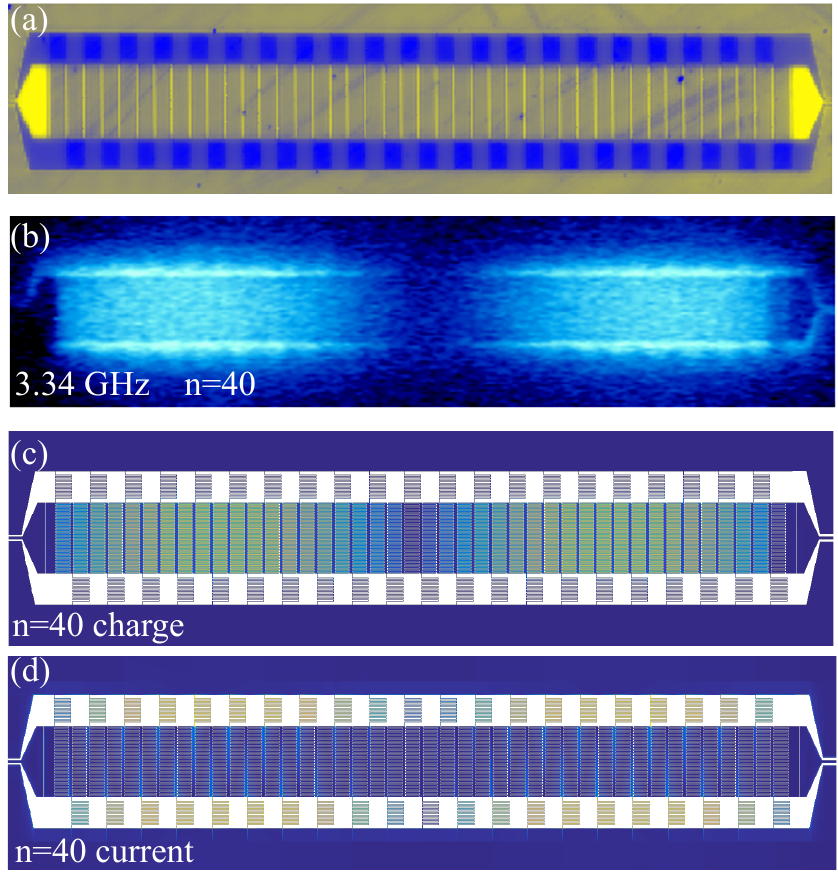}
  \caption{(LSM photoresponse compared with Sonnet current and charge density simulation: (a)~Reflectivity map from LSM showing details of metamaterial layout in the imaged area. (b)~2D distribution of  LSM photoresponse $\mathscr{R}(x,y, 3.44\,{\rm GHz})$ for the  $n=40$ mode. Bright (dark) regions correspond to large (small) PR signal. Sonnet simulations of the  $n=40$ mode (c)~charge density and (d)~current density.
\label{fig:ZoomedIn}}
\end{figure}
\begin{figure*}[htb]
\centering
\includegraphics[width=7in]{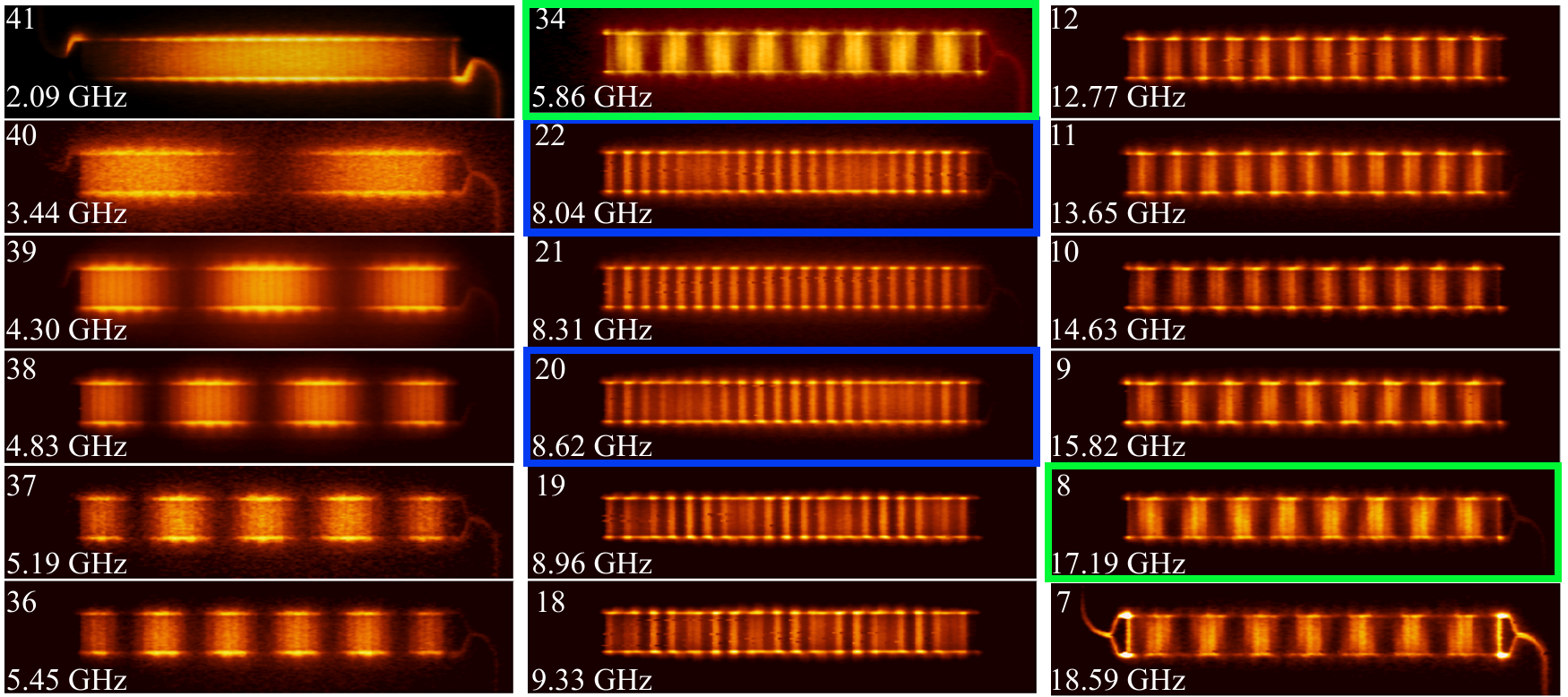}
  \caption{Array of LSM images of metamaterial for different modes, labeled by mode number and frequency. Bright (dark) regions correspond to large (small) PR signal. Amplitudes of PR signal are normalized for best contrast. Green boxes indicate the pair of $n=34$ and $n=8$ modes, which both have 8 antinodes. Blue boxes indicate the pair of $n=22$ and $n=20$ modes, which both have 20 antinodes.
\label{fig:LSM}}
\end{figure*}


We then perform 2D scans of $\mathscr{R}(x,y,f)$ while exciting one of the metamaterial modes. In Fig.~\ref{fig:ZoomedIn}(b), we present such a 2D scan for the $n=40$ mode at 3.34~GHz, which is the next-to-lowest resonant frequency in the device. The LSM image exhibits a clear standing-wave pattern with two antinodes along the length of the transmission line. We perform numerical electromagnetic simulations of our metamaterial layout using Sonnet, as aredescribed in detail in Sec.~\ref{sec:Sonnet}. Figures~\ref{fig:ZoomedIn}(c, d) contain plots of the simulated charge density and current density in the metamaterial, also excited on the $n=40$ mode. 

With the application of LSM imaging to superconducting structures, the interaction of the laser spot with the superconductor can produce a bolometric response, with both inductive and resistive components, due to local heating of the superconductor by the locally  absorbed laser power. In addition, there can be a non-thermal response due to the generation of nonequilibrium quasiparticles in the superconductor from the deposited pair-breaking laser energy. All three types of photoresponse signals are proportional to the square of the local microwave current density in the superconductor, $J_{MW}(x, y)$ \cite{zhuravel2010effect}. In our LSM imaging experiments on the LHTLs, we observe a significant photoresponse in regions where we expect large microwave currents, in particular, in the vicinity of the inductors near the current antinodes in the standing-wave patterns. The frequency dependence of the photoresponse on our LHTLs [Fig.~\ref{fig:LSMS21}(c)] and its correlation with $|S_{21}(f)|$ [Fig.~\ref{fig:LSMS21}(b)] is most consistent with the nonthermal LSM response mechanism \cite{zhuravel2006}. However, we observe an even larger photoresponse in regions where we expect large microwave voltages, around the capacitors near the standing-wave antinodes for the charge density. This appears to be caused by an enhancement in the loss in the capacitors due to photoinduced carriers in the Si substrate in between the fingers of the interdigitated capacitors. This mechanism results in the largest photoresponse in regions with large microwave electric fields. We are not aware of prior LSM imaging experiments of superconducting interdigitated capacitors on Si. Thus, the nature of our sample layout and substrate material provide us with a photoresponse image of both the microwave current and charge-density distributions in our LHTLs.

In the LSM image of Fig.~\ref{fig:ZoomedIn}(b) for the capacitors in the LHTL near an antinode, we observe that the photoresponse is highest at the edges of the capacitor where the inductors are connected. This effect can be attributed to the enhanced reflectivity of the laser signal from the regions covered with Nb compared with the bare Si regions beyond the capacitor edge. When the laser spot straddles the edge of the capacitor and the bare Si beyond, the excess absorbed energy in the Si enhances the photoresponse signal.

We repeat the LSM imaging for all of the modes in our LHTL below 20~GHz. We present an array of many of these in Fig.~\ref{fig:LSM}. The various features in these images are discussed in the next section.

\section{Analysis of LSM images}
\label{sec:analysis of LSM}
\begin{figure}[hb]
\centering
\includegraphics[width=3.35in]{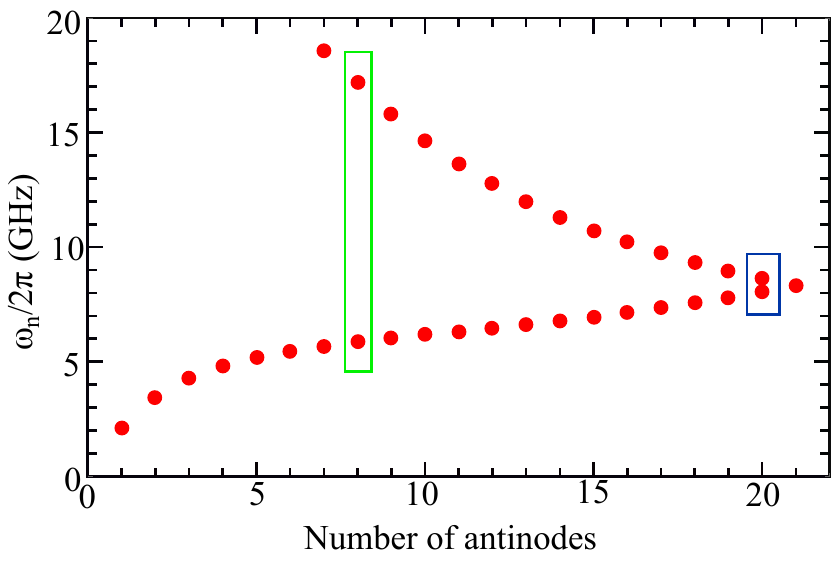}
  \caption{Plot of mode frequency vs. number of antinodes in the photoresponse from the corresponding LSM images. Colored boxes indicate the partner modes shown in Fig.~\ref{fig:LSM}.
\label{fig:freq-v-antinodes}}
\end{figure}

From the LSM images of the metamaterial resonator (Fig.~\ref{fig:LSM}), we observe a stepwise change in the number of antinodes in the standing-wave profile between neighboring modes. To understand these images, we  analyze them separately for low and high mode numbers. For the higher frequency modes, $n\le21$, it is clear that the wave number is decreasing with increasing frequency, consistent with the left-handed dispersion relation from Eq.~(\ref{eq:Caloz-CRLH-disp}). However, for the lower-frequency modes, from $n=$ 41 to 21, the images show a standing-wave pattern with an increasing number of antinodes for increasing mode frequency, a typical behavior of right-handed media. A plot of the mode frequency vs. the number of antinodes is shown in Fig.~\ref{fig:freq-v-antinodes}. For a given number of antinodes we see two standing-wave profiles with different frequencies, one at the low-frequency part of the band, and  one at high frequency. These partner modes can be observed at least up to almost 20~GHz. Figures~\ref{fig:LSM} and~\ref{fig:freq-v-antinodes} contain colored boxes indicating two such pairs of modes.

\begin{figure}[hb]
\centering
\includegraphics[width=3.35in]{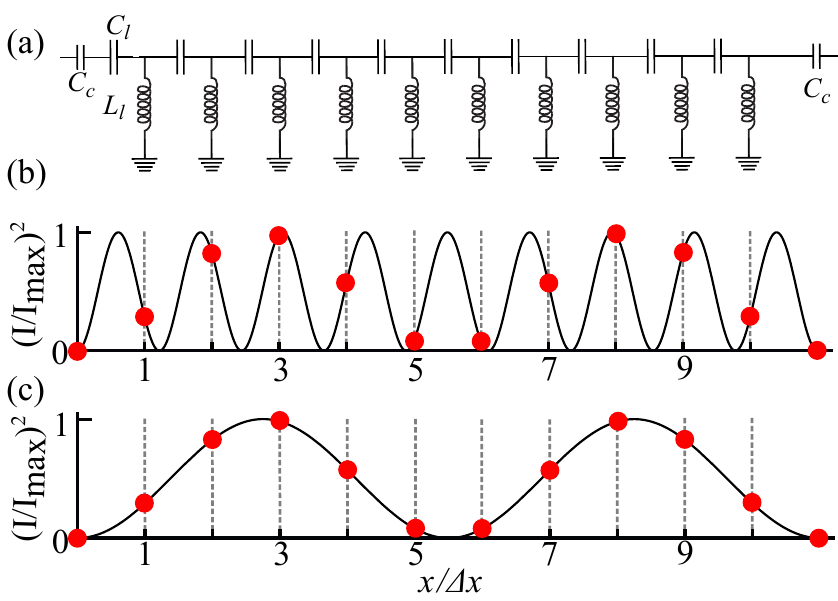}
  \caption{(a) Circuit schematics for ideal ten-cell metamaterial resonator. Waveform for the square of standing-wave current for hypothetical continuous LHTL for mode (b) $n=8$, (c) $n=2$, including red circles corresponding to location of each unit cell. 
\label{fig:undersampe-schem}}
\end{figure}

We can understand this behavior as an undersampling effect due to both the discrete lumped-element nature of our metamaterial transmission line and due to the LSM measurement itself. To demonstrate this, we consider a simplified model of an ideal 10-cell metamaterial resonator with no stray reactances, sketched in Fig.~\ref{fig:undersampe-schem}(a). 
For now, we focus on the microwave currents in the inductors, but the same analysis also applies to the voltage across the capacitors. Our open-ended boundary conditions imply current standing-wave patterns of the form $\sin\left(k_n x\right)$, with $k_n = n\pi/L$. However, the LSM photoresponse is proportional to the square of the local current density, so in Fig.~\ref{fig:undersampe-schem}(b, c) we plot the square of this continuous waveform. The figure contains the continuous waveforms for a low-frequency mode and its partner high-frequency mode $N-p$ and $p$, which are clearly different. However, the current-square values at the unit-cell locations, indicated by red dots in the figure, appear the same. This undersampling effect is consistent with the LSM images and the plot of $\omega_n$ vs. the number of antinodes in Fig.~\ref{fig:freq-v-antinodes} if we extend the example circuit in Fig.~\ref{fig:undersampe-schem} to 42 cells.

%

\begin{figure}[htb]
\centering
\includegraphics[width=3.35in]{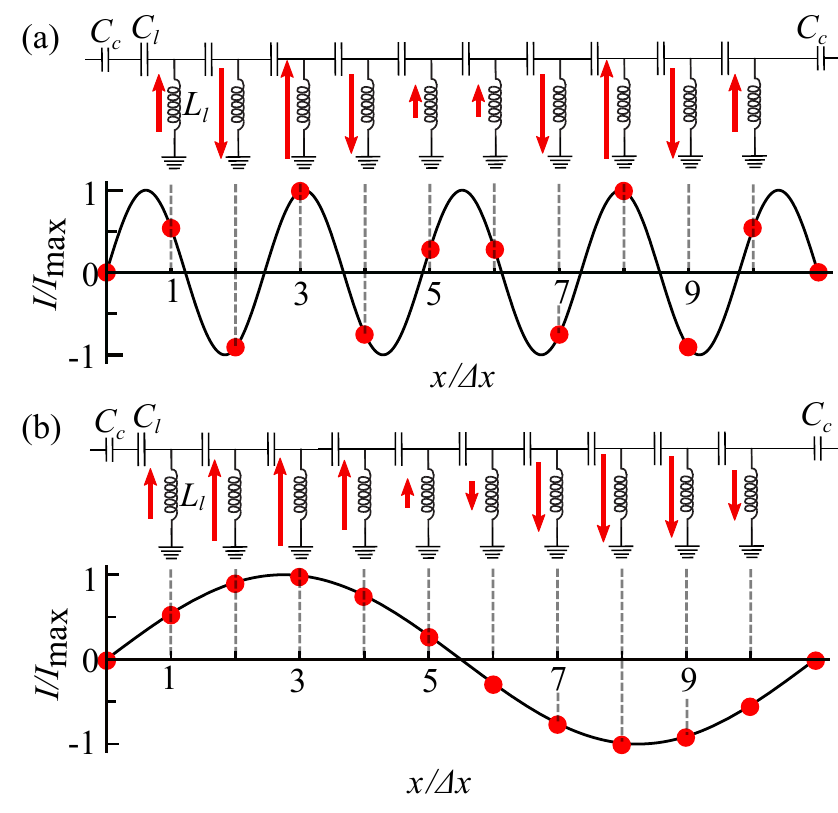}
  \caption{Circuit schematics for ideal ten-cell metamaterial resonators with arrows indicating the relative sense of currents flowing in each unit cell along with waveform for standing-wave current for hypothetical continuous LHTL. Red circles correspond to current at location of each unit cell for mode (a) $n=8$, (b) $n=2$. 
\label{fig:current-mag-schem}}
\end{figure}

Besides a comparison with the LSM images, we can also use this approach to understand more details about the nature of the mode patterns in the metamaterial resonators. If we again consider the ideal ten-unit-cell device, but plot the local current vs. position rather than the square of the current, we can again compare the continuous waveform for the standing-wave current along with the corresponding discrete values at each of the unit cells (Fig.~\ref{fig:current-mag-schem}). This analysis indicates that the low-frequency modes, with mode number $n>N/2$, should have currents that alternate in sign between adjacent unit cells. As the LSM is only sensitive to the square of the local current density, we are unable to resolve these relative phases.


\begin{figure}[htb]
\centering
\includegraphics[width=3.35in]{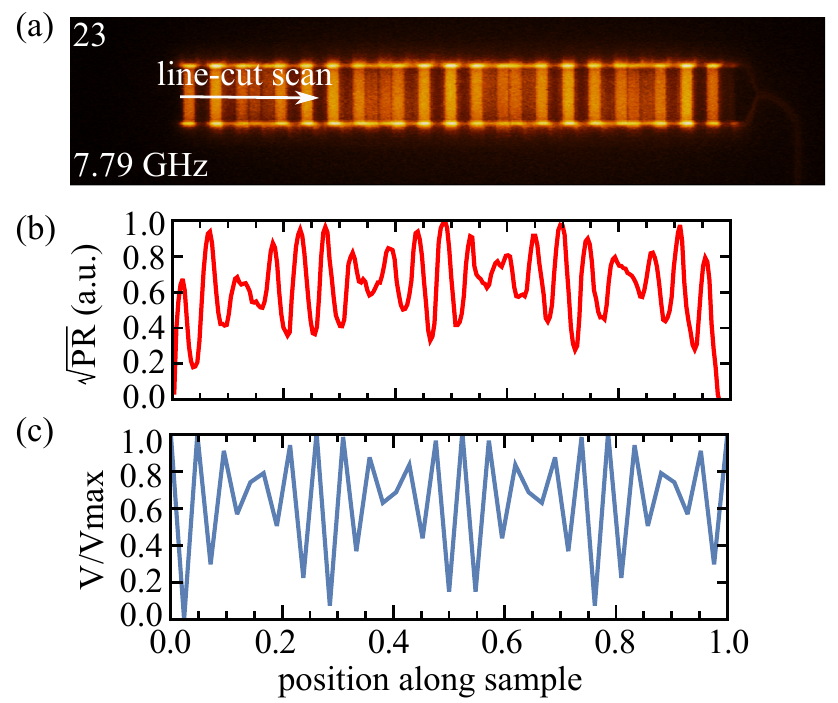}
  \caption{(a) LSM photoresponse for mode 23 with the arrow indicating the location of subsequent line cut. Bright (dark) regions correspond to large (small) PR signal. (b) Plot of the square root of LSM photoresponse signal along line cut. (c) Standing-wave pattern of voltage across capacitors computed with Eq.~(\ref{eq:V_Clm}) for mode 23 of a 42-cell metamaterial resonator.
\label{fig:LSMBeating}}
\end{figure}

This undersampling effect is also the origin for a beating pattern that is particularly prominent for modes near $N/2$. Figure~\ref{fig:LSMBeating}(a) shows the LSM image for mode $n=23$, which exhibits such a beating pattern, along with a line-cut  of the square root of the photoresponse along the center of the image in Fig.~\ref{fig:LSMBeating}(b). Appendix~\ref{appendix:phase} details the calculation of current and voltage values at the circuit nodes for an ideal LHTL resonator. Although nonideal lumped-element effects may be important, especially at short wavelengths, for now we assume that the currents and voltages are constant within each unit cell. The calculated voltage across the capacitors from Eq~(\ref{eq:V_Clm}), shown in Fig.~\ref{fig:LSMBeating}(c), displays reasonable agreement with the measurement. 


\section{Numerical simulations of metamaterials}
\label{sec:Sonnet}
\begin{figure}[htb]
\centering
\includegraphics[width=3.35in]{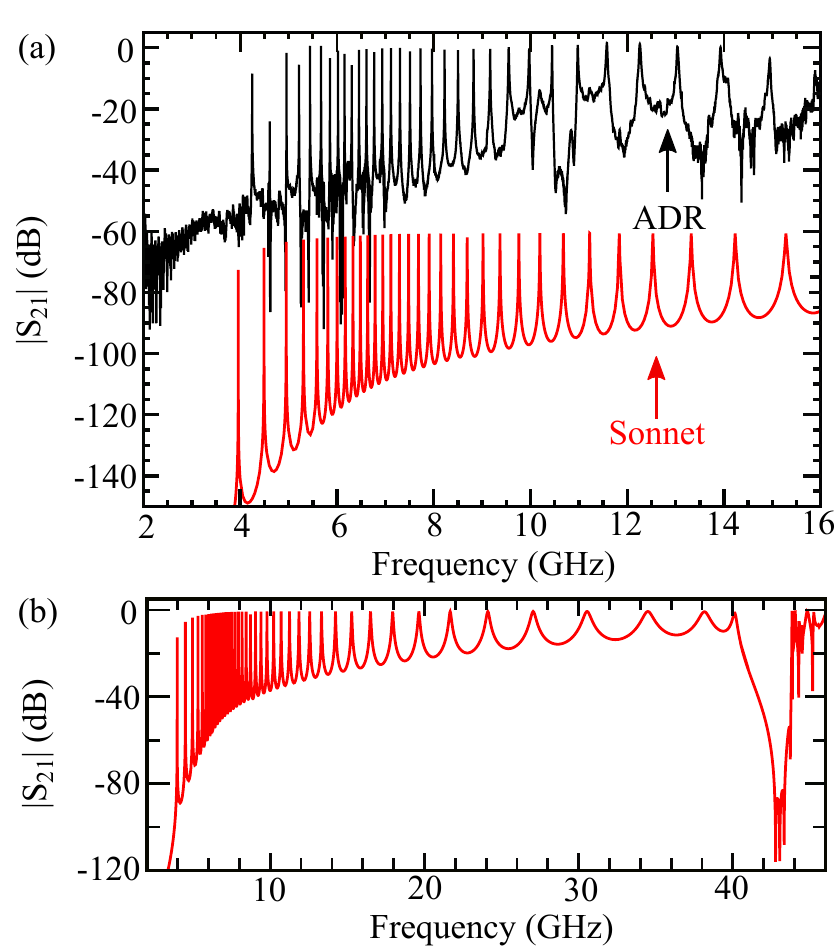}
  \caption{(a)~$|S_{21}(f)|$ from Sonnet simulation (red) offset by 60 dB compared with measured LHTL spectrum from ADR at 65~mK~(black). (b)~$|S_{21}(f)|$ from Sonnet simulation plotted out to $45\,{\rm GHz}$ showing gap beyond $n=0$ mode near 40 GHz.
\label{fig:sonnet}}
\end{figure}

\begin{figure}
\centering
\includegraphics[width=3.35in]{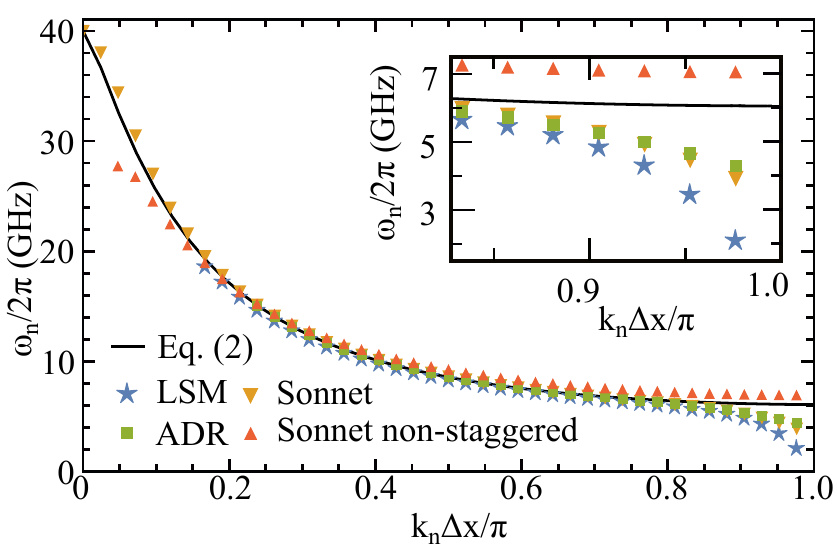}
  \caption{Plot of mode frequency vs. normalized waven umber for ADR measurements, LSM images, Sonnet simulations, including both staggered and non-staggered inductor configurations, as discussed in the text; the solid black curve corresponds to dispersion relation of Eq.~(\ref{eq:Caloz-CRLH-disp}). The inset shows a closeup of behavior near $\omega_{IR}/2\pi$.
\label{fig:Freq-mode-all}}
\end{figure}

We simulate the microwave properties of our metamaterial structures using the Sonnet electromagnetic field solver~\cite{plourde2015superconducting}. We compute the $S_{21}$ transmission spectrum, as well as the current-density and charge-density distributions, near the metamaterial resonance frequencies. 
In Fig.~\ref{fig:sonnet}(a) we compare the simulated $|S_{21}(f)|$ from Sonnet with our measured spectrum that is presented earlier in Fig.~\ref{fig:T-dep}. The simulated curve is in reasonable qualitative agreement with the experimentally measured spectrum, although there are notable deviations, particularly at the lowest frequency modes. Both the Sonnet and experimental spectra exhibit the highest density of modes around 6~GHz, in contrast to the prediction from Eq.~(\ref{eq:Caloz-CRLH-disp}) that the mode density should be highest just above $\omega_{IR}$ where the band for the ideal circuit is flat. We also observe good agreement between the resonance frequencies from the simulated spectrum, theoretical dispersion relation from Eq.~(\ref{eq:Caloz-CRLH-disp}), and the experimentally measured values from both the ADR and LSM over much of the frequency range of our measurements (Fig.~\ref{fig:Freq-mode-all}). However, there are discrepancies between these various methods for the modes near the low-frequency end of the spectrum. 

Modes just above $\omega_{IR}$ correspond to the highest wave number and thus the fastest spatial oscillations, where the wavelength is approaching twice the unit cell size. Thus, deviations from ideal lumped-element behavior are most prominent in this regime. One such deviation arises from our choice of layout. Due to the constraints of our microwave chip holder, we chose a metamaterial design with staggered inductors, where inductors on the odd-numbered unit cells terminate on one side of the ground plane, while those on the even-numbered cells connect to the other side. This permitted us to fit more unit cells along the length of our metamaterial resonator without having the inductors of neighboring cells interfering with each other. However, this arrangement for the inductors allows for the possibility of non-lumped-element behavior impacting the frequencies of the high wave-number modes. To explore this further, we  implement a metamaterial layout in Sonnet with the same $L_l$ and $C_l$ values as in our other devices, but with all of the inductors on one side of the transmission line. This requires that the simulated structure is 10\% longer than our measured metamaterial resonators. From the simulations of the metamaterial with nonstaggered inductors, we observe that the mode frequencies are consistent with the prediction from Eq.~(\ref{eq:Caloz-CRLH-disp}), even up to the highest wave-number mode (Fig.~\ref{fig:Freq-mode-all}). Further details of the inductor staggering characterization are discussed in detail in Appendix~\ref{sec:staggered}. 

Besides the frequency shifts arising from the staggered inductors, another factor that impacts the frequencies of the high wave-number modes is the quality of the microwave grounding of the chip. This is evident in Fig.~\ref{fig:Freq-mode-all}, where the mode frequencies  of the high $-n$  resonances measured on the LSM, where it is especially challenging to maintain a good microwave ground across the chip while permitting the LSM imaging process, are even further shifted from the dispersion relation of Eq.~(\ref{eq:Caloz-CRLH-disp}) compared to the ADR measurements and the Sonnet values. We simulate the effects of degraded ground connections for the metamaterial using Sonnet and observed that this indeed results in downward shifts of the resonance frequencies for the modes with the highest wavenumber. Appendix~\ref{appendix:ImpGround} contains a more detailed discussion of the effects of ground quality on the mode spectrum. 

In addition to studying the effects of different inductor layouts and imperfect grounding, we can also use Sonnet to explore the spectrum of our metamaterial resonators for frequencies beyond the range of our experimental hardware. In Fig.~\ref{fig:sonnet}(b), we extend the Sonnet simulation of our metamaterial up to 46~GHz, where it is clear the mode spacing continues to increase and the linewidths become progressively larger. In addition, beyond the $n=0$ mode, we observe a gap of roughly 4~GHz in the simulated spectrum. This is consistent with the theoretical prediction from Eq.~(\ref{eq:Caloz-CRLH-disp}) for a metamaterial with different self-resonance frequencies for the inductor and capacitor elements~\cite{caloz2004novel}.
We   run  separate  simulations  in  Sonnet of  an  individual  inductor  and  capacitor  from  our  metamaterial  design  and  obtained  self-resonance  frequencies of approximately $50$~GHz  for  the  capacitor  and approximately $60$~GHz  for  the inductor,  close  to  the  location  of  the  gap  in  the  simulated LHTL resonator spectrum  of  40-44~GHz. The  simulation  of  individual capacitors and inductors may not capture the actual grounding environment or  loading effects  that  arise  when  the  elements  are  built  into  a  metamaterial,  thus leading to differences in the simulated self-resonance frequencies.  Nonetheless, the simulations give qualitative agreement between the capacitor and inductor self-resonances and the gap in the metamaterial spectrum. If we take the self-resonance frequencies from the gap frequencies, they correspond to stray reactances of $L_r = 59.5$~pH and $C_r = 21.8$~fF. For the Sonnet simulations of the alternate metamaterial layout with non-staggered inductors discussed earlier, we observe in Fig.~\ref{fig:Freq-mode-all} that the high-frequency, low$-n$ modes are shifted to somewhat lower frequencies than those of the layout for our measured device, consistent with a reduction in the inductor self-resonance frequency due to the different geometry.

We note that the simulation of the transmission spectra for our metamaterial design is rather time consuming. We use a mesh size of $1\,\mu{\rm m}$ to capture the smallest feature size in our layout for the interdigitated capacitors and meander-line inductors. These relatively small feature sizes, combined with the large spatial extent of our metamaterial resonator, pose a significant challenge for a numerical field solver.  Using a small-scale computing cluster with 10 slave machines, each with 16 cores and 64 GB of memory, a simulation of a complete spectrum, as in Fig.~\ref{fig:sonnet}, takes approximately 3.5 weeks. However, once the spectrum computation is completed, we are able to run simulations of the current-density and charge-density distributions at particular frequencies, with scans at all 42 modes requiring about 2 days to complete. We present examples of such distributions in other sections of the manuscript, including Sec.~\ref{sec:LSM imaging of mode structure} and Appendix~\ref{appendix:phase}.

\section{Coupling and Internal Losses}
\label{sec:loss}
The quality factor of the metamaterial resonances can be determined by standard fits to the trajectory of $S_{21}(f)$ in the complex plane, in the same way as with a resonator formed from either a conventional transmission line or lumped elements \cite{Pappas2011,sage2011study}. For a dip-style reflection measurement of a resonator coupled to a feedline where the baseline level self-calibrates full transmission, the fitting process allows for a simultaneous extraction of both internal losses $Q_i^{-1}$ and losses due to coupling to dissipation in external circuitry $Q_c^{-1}$. However, for peak-style transmission resonances, as with the metamaterial measurements presented here, one requires a separate baseline calibration in order to separate the values of $Q_i$ and $Q_c$ from the fit. Such baseline measurements inevitably introduce an extra source of variation since they must be performed on a separate cooldown with a different device that may have slight differences in the bonding and packaging.

\begin{figure}
\center
\includegraphics[width=3.35 in]{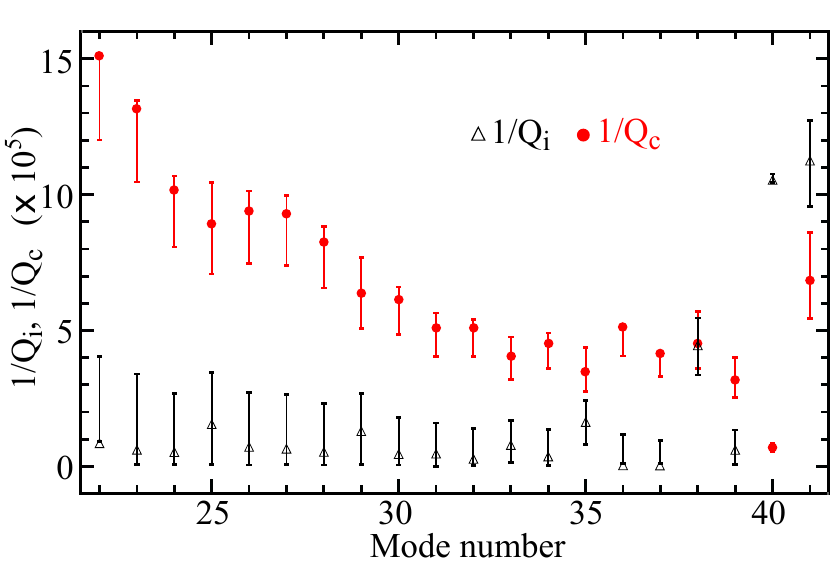}
\caption{Comparison of internal loss and coupling loss extracted from the measured $S_{21}(f)$.}
\label{fig:MeasuredQ}
\end{figure}

Following a separate baseline measurement on our ADR, we obtain the normalized spectrum for our metamaterial resonator in Fig.~\ref{fig:T-dep}. We then fit each resonance and extract $Q_c$ and $Q_i$, allowing us to plot the internal and external losses as a function of mode number $n$ in Fig.~\ref{fig:MeasuredQ}. It is clear that most of the modes are strongly overcoupled, with $Q_c^{-1} \gg Q_i^{-1}$, consistent with the majority of the resonance peaks in the spectrum of Fig.~\ref{fig:T-dep} reaching close to full transmission. The error bars are determined by an estimate of the uncertainty in the baseline transmission level from variations in $|S_{21}|$ for cooldowns of multiple nonmetamaterial devices exhibiting broadband full transmission with the same wiring as on our metamaterial measurements. We take this variation to be $\pm 2\,{\rm dB}$, taking care to ensure that the normalized $|S_{21}|$ never exceeds unity.

From Fig.~\ref{fig:MeasuredQ}, the $Q_i^{-1}$ values for most of the modes are of the order of $10^{-5}$, consistent with dielectric loss in the interdigitated capacitors that make up the metamaterial \cite{Oconnell08, martinis2005decoherence}. However, a few of the lowest frequency, highest $n$ modes have an order of magnitude higher internal loss, which may be related to the short wavelengths for these modes or non-lumped-element effects due to the staggering of the inductors for this particular metamaterial layout.
\begin{figure}[b]
\centering
\includegraphics[width=3.35in]{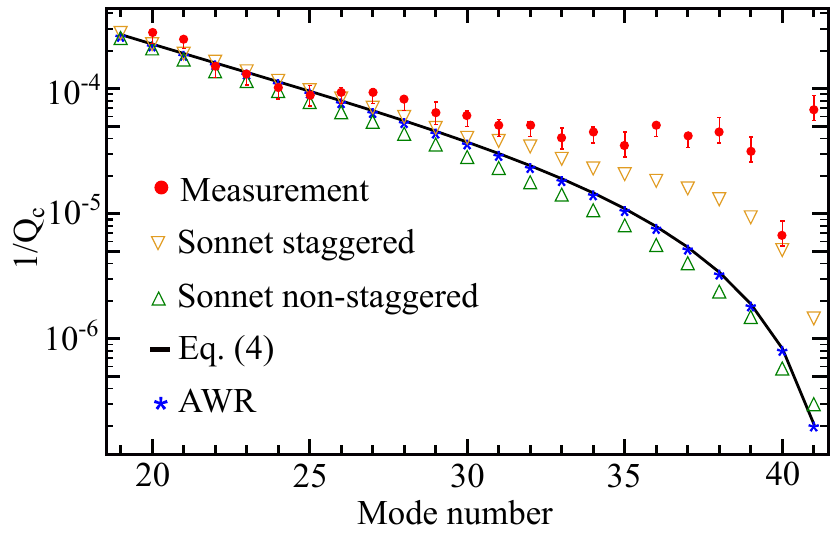}
  \caption{Coupling losses for different mode numbers extracted from experimentally measured resonances on ADR at 65~mK, AWR simulation of ideal lumped-element LHTL resonator, and Sonnet simulations of both staggered and nonstaggered inductor configurations. the solid curve corresponds to Eq.~(\ref{eq:QcLHL1}). 
\label{fig:losses}}
\end{figure}

The $Q_c^{-1}$ values generally decrease for larger $n$ modes, although the highest $n$ mode deviates from this trend, perhaps for the same reasons as the $Q_i^{-1}$ behavior of the high-$n$ modes described above. For a conventional, distributed transmission-line resonator, formed from perhaps a coplanar waveguide, the mode-number dependence of the coupling loss is quite different: $Q_c^{-1} \propto n$~\cite{Goppl08}. In Appendix~\ref{sec:couplingloss}, we derive expressions for $Q_c$ for resonators formed from discrete lumped-element transmission lines, with both right-handed (which we have not seen treated previously in the literature) and left-handed dispersion relations. For the simplest case of an ideal LHTL resonator, we obtain

\begin{equation}
Q_c^{LHTL}(n)^{-1}=\frac{C_c^2\cos^2\left(\frac{n\pi}{2N}\right)}{2NC_l^2\sin^3\left(\frac{n\pi}{2N}\right)}\label{eq:QcLHL1}.
\end{equation}


In Fig. \ref{fig:losses}, we plot $Q_c^{-1}(n)$ from Eq.~(\ref{eq:QcLHL1}), along with values from a numerical simulation of an ideal lumped-element LHTL resonator using AWR Microwave Office showing nearly perfect agreement. The $Q_c^{-1}$ values extracted from our measured resonances agree reasonably well with Eq.~(\ref{eq:QcLHL1}) for lower values of $n$, but begin to deviate significantly for higher $n$ modes starting around $n \sim 30$. From fits to extract $Q_c$ from the resonances in our Sonnet simulations, the $Q_c^{-1}$ values for the nonstaggered inductor configuration follow the dependence of Eq.~(\ref{eq:QcLHL1}) rather closely. However, the values from the Sonnet simulations of the same staggered-inductor layout as with our experimental device also follow the trend of excess coupling loss for the highest $n$ modes. This effect is likely a consequence of the short wavelengths for these high-$n$ modes, which are strongly influenced by the layout of the inductors, as well as the integrity of the ground plane across the chip (Appendix~\ref{appendix:ImpGround}).

\section{Conclusions}
\label{sec:conclusions}
In conclusion, we demonstrate superconducting thin-film metamaterial resonators that produce dense mode spectra with left-handed dispersion in a frequency range that is compatible with the integration of superconducting qubits in a cQED architecture. We perform LSM imaging of the microwave field distributions on one of these devices and compare this with both analytical modeling of the circuit as well as numerical simulations that allow us to predict the spectrum of new metamaterial designs. While our current minimum mode spacing is 147~MHz, this could be reduced by increasing the number of unit cells $N$ or by intentionally reducing the self-resonance of the capacitive or inductive elements, perhaps by using a high kinetic inductance film for the capacitor fingers or by designing intentional capacitive shunts on the inductors. Metamaterial resonators with substantial nonlinearity and frequency tunability could also be developed by incorporating Josephson elements in place of the linear inductors in our present design.

Our circuit layout allows for straightforward extensions to future devices incorporating superconducting qubits that couple to the metamaterial resonances. This would potentially open up a new regime for analog quantum simulation with the spectrum of photonic modes tailored by the circuit parameters making up the metamaterial. In addition to superconducting qubits, our metamaterial design is also compatible with coupling to nanomechanical devices \cite{rouxinol2016}, quantum-dot qubits \cite{Frey2012,Petersson2012}, or even qubits based on hyperfine transitions in trapped ions or Rydberg atoms in hybrid superconducting/atomic systems \cite{Pritchard2014}.

\section*{Acknowledgments}
This work is supported by the U.S. Government under ARO Grants No. W911NF-14-1-0080 andNo.  W911NF-15-1-0248. Device fabrication is performed at the Cornell NanoScale Facility, a member of the National Nanotechnology Infrastructure Network, which is supported by the National Science Foundation (Grant No. NNCI-1542081). A.P.Z. acknowledges support from Volkswagen Foundation (Grant No. 90284).
B.G.T. acknowledges support from FAPESC, CNPq and the National Institute for Science and Technology - Quantum Information. A.V.U. acknowledges partial support from the German Science Foundation (Grant No. US18/15-1), the Russian Science Foundation (Grant No. 16-12-00095), and the Ministry of Education and Science of Russian Federation in the framework of the Increase Competitiveness Program of the National University of Science and Technology MISIS (Grant No. K2-2017-081).

\appendix
\section{Derivation of dispersion relation}
\label{appendix:Derivation of dispersion relation}
\begin{figure}[htb]
\center
\includegraphics[width=3.35in]{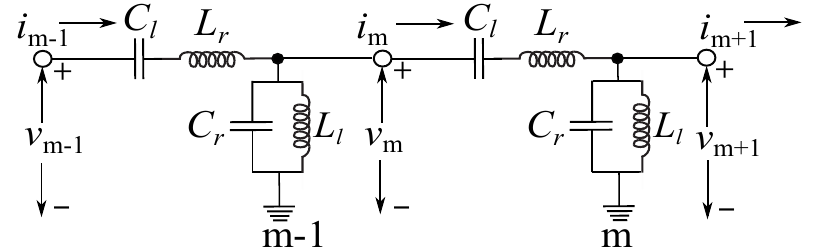}
   \caption{Voltage and current in a unit cell of a general discrete transmission line. 
\label{fig:unitcell}}
\end{figure}
For a general discrete transmission line, such as the circuit sketched in Fig.~\ref{fig:meta-schem}(b), we can study the current and voltage
supported by such a line by focusing on two adjacent unit cells, as in Fig.~\ref{fig:unitcell}, and applying Kirchhoff's laws to obtain the following relations


\begin{align}
v_m-v_{m+1}=&i_m\left(i\omega L_r+\frac{1}{i\omega C_l}\right)\label{eqn:A1}\\
i_{m-1}-i_m=&v_m\left(i\omega C_r+\frac{1}{i\omega L_l}\right)\label{eqn:A2}\\
v_{m-1}-v_m=&i_{m-1}\left(i\omega L_r+\frac{1}{i\omega C_l}\right)\label{eqn:A3}\\
i_m-i_{m+1}=&v_{m+1} \left(i\omega C_r+\frac{1}{i\omega L_l}\right)\label{eqn:A4}.
\end{align}

\noindent By defining the admittance $Y=i\omega C_r+1/i\omega L_l$ and impedance $Z=i\omega L_r+1/i\omega C_l$, we can reduce these to two expressions
\begin{align}
v_m(2+ZY)=&v_{m-1}+v_{m+1}\label{eqn:A5}\\
i_m(2+ZY)=&i_{m-1}+i_{m+1}\label{eqn:A6}.
\end{align}

\noindent Assuming a plane wave solution for propagation through the transmission line, the voltage and current for cell $m$ can be written as
\begin{align}
v_m=&V_0^{+}e^{-ik m\Delta x}+V_0^{-}e^{ik m\Delta x\label{eqn:A7}}\\
i_m=&I_0^{+}e^{-ik m\Delta x}+I_0^{-}e^{ik m\Delta x}\label{eqn:A8},
\end{align}

\noindent where $\Delta x$ is the unit cell length, and $V_0^+ (V_0^-)$ and $I_0^+ (I_0^-)$ are the amplitudes of the forward (reverse) propagating voltage and current, respectively. Combining  Eqs.~(\ref{eqn:A7})-(\ref{eqn:A8}) with Eqs.~(\ref{eqn:A5})-(\ref{eqn:A6}), we obtain
\begin{equation}
[V_0^+e^{-ik m\Delta x}+V_0^-e^{ik m\Delta x}][2\cos\left(k \Delta x\right)-(2+ZY)]=0.
\label{eq:appendixAequ9}
\end{equation}
This expression must be satisfied for all values of $k m\Delta x$, therefore
\begin{equation}
2\cos\left(k \Delta x\right)=(2+ZY)\label{eq:A10}.
\end{equation}
This relationship between $k\Delta x$ and $ZY$ leads to the dispersion relation for the transmission line
\begin{equation}
\begin{split}
k(\omega) =& \frac{1}{\Delta x}\times\\
&\cos^{-1}\left[1-\frac{1}{2}\left(\omega L_r-\frac{1}{\omega C_l} \right) \left(\omega C_r-\frac{1}{\omega L_l} \right) \right],
\end{split}\label{eq:A11}
\end{equation}
matching Eq.~(\ref{eq:Caloz-CRLH-disp}) in the main text. For the case of a purely right-handed line (i.e., $1/C_l$=0 and $1/L_l$=0), using standard trigonometric relationships combined with Eq.~(\ref{eq:A10}), inverting Eq.~(\ref{eq:A11}) yields the dispersion relation for a right-handed line:
\begin{equation}
\omega _R(k)=\frac{2}{\sqrt[]{L_r C_r}}\sin\left(\frac{k\Delta x}{2}\right)\label{eq:RHdispersion}.
\end{equation}
On the other hand, if the line is purely left-handed (i.e., $L_r$=0, $C_r$=0), the admittance becomes $1/i\omega L_l$ and the impedance becomes $1/i\omega C_l$. Again, inverting Eq.~(\ref{eq:A11}) with these conditions leads to the left-handed dispersion relation for an ideal LHTL
\begin{equation}
\omega _L(k)={\frac{1}{2\sqrt{L_l C_l}}\frac{1}{\sin\left(\frac{k\Delta x}{2}\right)}}\label{eq:LHdispersion}.
\end{equation}


\section{Impedance of general terminated discrete transmission line}
\label{sec:Impedance of general terminated discrete transmission line}
In order to treat a resonator formed from a discrete transmission line, we need to continue our analysis from Appendix~\ref{appendix:Derivation of dispersion relation} to derive the impedance for a discrete transmission line with a general terminating load. Before treating the addition of the termination, we revisit Eqs.~(\ref{eqn:A7}) and (\ref{eqn:A8}) and substitute into Eqs.~(\ref{eqn:A1})-(\ref{eqn:A4}) to obtain
\begin{align}
I_0^+=&2ie^{-ik\Delta x/2}\sin{\left(\frac{k\Delta x}{2}\right)} \frac{V_0^+}{Z}\label{eq:A14}\\
I_0^-=&-2ie^{ik\Delta x/2}\sin\left(\frac{k\Delta x}{2}\right) \frac{V_0^-}{Z}\label{eq:A15}.
\end{align}
Following these expressions, we define the characteristic impedance of the line to be $Z_0=Z/2i\sin{\left(k\Delta x/2\right)}$ and we can express the current in cell $m$ in terms of $Z_0$:
\begin{equation}
i_m=\frac{V_0^+}{Z_0}e^{-ik(m+\frac{1}{2})\Delta x}-\frac{V_0^-}{Z_0}e^{ik(m+\frac{1}{2})\Delta x}.
\end{equation}

Next, we consider a finite length discrete transmission line with $N$ unit cells and the last cell on the right being terminated by a load impedance $Z_l$, which forms an extra cell. To simplify the calculation, we choose the cell of the terminating load to be $m=0$. Introducing the reflection coefficient $\Gamma = V_0^-/V_0^+$, the voltage and current at cell $m$ can then be expressed in terms of $\Gamma$ as
\begin{align}
v_m=&V_0^{+}e^{-ik m\Delta x}+\Gamma V_0^{+}e^{ik m\Delta x\label{eq:VmwithGamma}}\\
i_m=&\frac{V_0^{+}}{Z_0}e^{-ik(m+\frac{1}{2})\Delta x}-\Gamma\frac{V_0^{+}}{Z_0}e^{ik(m+\frac{1}{2})\Delta x}\label{eq:ImwithGamma}.
\end{align}

\noindent The impedance at $m=0$ can be found as
\begin{equation}
Z_{m=0}=\frac{v_0}{i_0}=Z_0\frac{1+\Gamma}{e^{\frac{-ik\Delta x}{2}}-\Gamma e^{\frac{ik\Delta x}{2}}}\label{eq:A19}.
\end{equation}
At cell $m=0$, we require that $Z_m=Z_l$. Thus, we have
\begin{equation}
\Gamma=\frac{Z_le^{\frac{-ik\Delta x}{2}}-Z_0}{Z_le^{\frac{ik\Delta x}{2}}+Z_0}\label{eq:A20},
\end{equation} 
which, it is important to note, is not equal to one if $Z_l \to \infty$, in contrast to a continuous transmission line.

With the reflection coefficient calculated, we can then calculate the impedance seen from the opposite end of the transmission line. Applying the convention of increasing cell number from left to right and our choice of $m=0$ for the terminating cell on the right end of the line, the first cell on the left is $m=-N$. Thus, based on Eqs.~(\ref{eq:VmwithGamma})-(\ref{eq:ImwithGamma}), we can write the voltage and current for the input cell at the left end as
\begin{align}
v_{-N}=&V_0^{+}e^{ik N\Delta x}+\Gamma V_0^{+}e^{-ik N\Delta x}\label{eqn:V-N}\\
i_{-N}=&\frac{V_0^+}{Z_0}e^{-ik(-N+\frac{1}{2})\Delta x}-\Gamma \frac{V_0^+}{Z_0}e^{ik(-N+\frac{1}{2})\Delta x}\label{eq:A22},
\end{align}
leading to the following expression for the impedance at the input cell
\begin{equation}
Z_{-N}=Z_0\frac{e^{ik N\Delta x}+\Gamma e^{-ik N\Delta x}}{e^{-ik(-N+\frac{1}{2})\Delta x}-\Gamma  e^{ik(-N+\frac{1}{2})\Delta x}}.\label{eqn:impedanceZN}
\end{equation}
This is a general result for a  lossless discrete transmission line. For example, it is straightforward to compute the input impedance for a loaded LHTL for a given mode number by applying the appropriate dispersion relation from Appendix~\ref{appendix:Derivation of dispersion relation} to Eq.~(\ref{eqn:impedanceZN}).

\section{Derivation of $S_{21}(\omega)$ for general discrete transmission line resonator}
\label{sec:S21discrete}
\begin{figure}
\center
\includegraphics[width=3.35in]{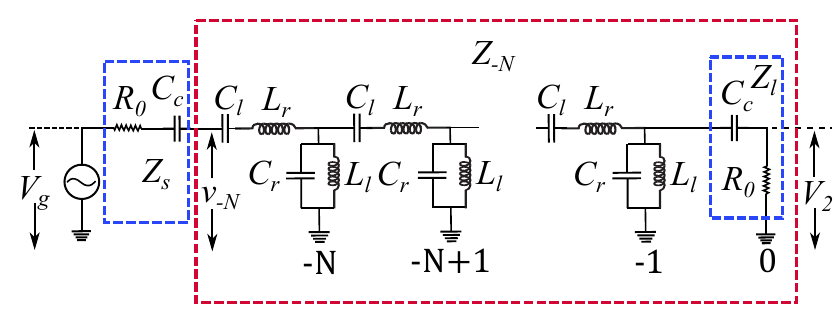}
   \caption{Circuit diagram for a $S_{21}$ measurement setup for an $N$-cell metamaterial transmission line resonator. We assume symmetric coupling so that the source impedance at the input end $Z_s$ and the load impedance at the terminated end $Z_l$ are equal. The total impedance of the LHTL resonator looking from the left end is $Z_{-N}$ given by Eq.~(\ref{eqn:impedanceZN}) and is marked in red dashed box. 
\label{fig:S21}}
\end{figure}

In order to derive a general expression for $S_{21}(\omega)$ through a discrete transmission line resonator, we take the terminating load to be determined by the output coupling capacitor $C_c$ and a resistive load $R_0$ such that the impedance of the terminating cell at $m=0$ is $Z_l=R_0+1/i\omega C_c$. We then add an input drive with source impedance $R_0$ and input coupling capacitance $C_c$ , such that $Z_s = Z_l$(Fig.~\ref{fig:S21}). For simplicity and to match our experimental configuration, we assume symmetric coupling. However, our analysis could easily be extended to the case of asymmetric coupling. For a source voltage $V_g$ and voltage $V_2$ across the output load $R_0$, the transmission $S_{21}$ is given by $2V_2/V_g$. Initially we consider the voltage at the first cell of the transmission line $v_{-N}$ and express this in terms of $V_g$
\begin{equation}
v_{-N}=V_g\frac{Z_{-N}}{Z_s+Z_{-N}}.
\end{equation}
Next we consider the voltage across $Z_l$ at the output end, which is $(Z_l/R_0)V_2$. This voltage should correspond to $V_0^+(1+\Gamma)$ and $V_0^+$ can be obtained from Eq.(\ref{eqn:V-N}). Thus, $V_2$ is given by
\begin{equation}
V_2=V_g\frac{Z_{-N}}{Z_s+Z_{-N}}\frac{R_0}{Z_l} \frac{1+\Gamma}{e^{ik N\Delta x}+\Gamma e^{-ik N\Delta x}}\label{eq:A25}.
\end{equation}
Finally, $S_{21}$ can then be written as
\begin{equation}
S_{21}=\frac{2Z_{-N}}{Z_s+Z_{-N}}\frac{R_0}{Z_l} \frac{1+\Gamma}{e^{ik N\Delta x}+\Gamma e^{-ik N\Delta x}}\label{eqn:S21}.
\end{equation}
$S_{21}$ can then be computed from Eq.~(\ref{eqn:S21}) by substituting Eq.~(\ref{eqn:impedanceZN}) for $Z_{-N}$ and Eq.~(\ref{eq:A20}) for $\Gamma$. Note that the frequency dependence of $S_{21}$ is determined by the dispersion relation $k(\omega)$ that one chooses.
\begin{figure}[htb]
\centering
\includegraphics[width=3.35in]{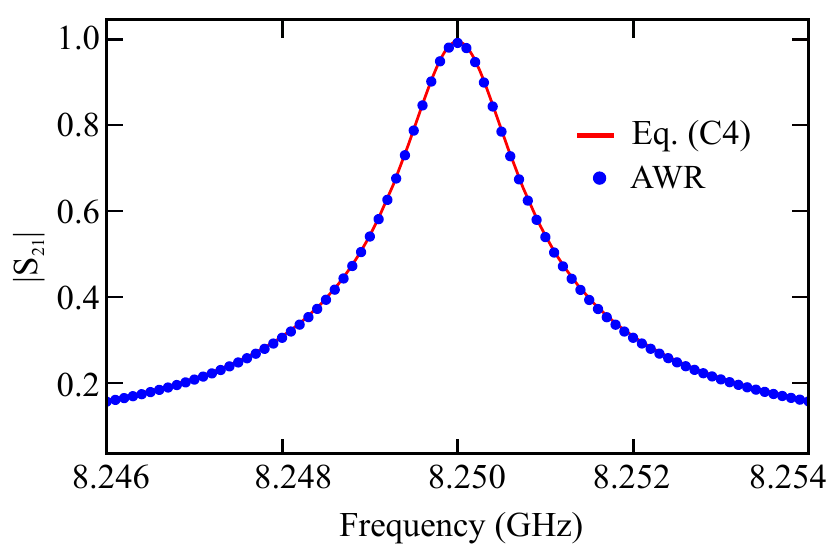}
  \caption{$S_{21}(f)$ calculated for the $n=23$ mode of a 42-cell LHRH metamaterial resonator using Eq.~(\ref{eqn:S21}) (solid red line) and simulated using AWR Microwave Office (blue points). Unit cell parameters are: $C_l=266$ fF, $L_{L}=0.6$ nH, $C_r=21.806$ fF and $L_r=0.595$ nH, chosen based on the discussion in Sec.~\ref{sec:Fab} and the Sonnet simulation result of the stray reactances. 
\label{fig:S21 plot comparison}}
\end{figure}
In Fig.~\ref{fig:S21 plot comparison}, we plot $S_{21}(\omega)$ computed from Eq.~(\ref{eqn:S21}) for the $n=23$ mode of a LHRH resonator with the parameters given in the caption. A numerical simulation of a circuit with these same parameters in AWR Microwave Office yields quite good agreement.


\section{Coupling quality factor for discrete transmission line resonators}
\label{sec:couplingloss}
\begin{figure}
\center
\includegraphics[width=3.35in]{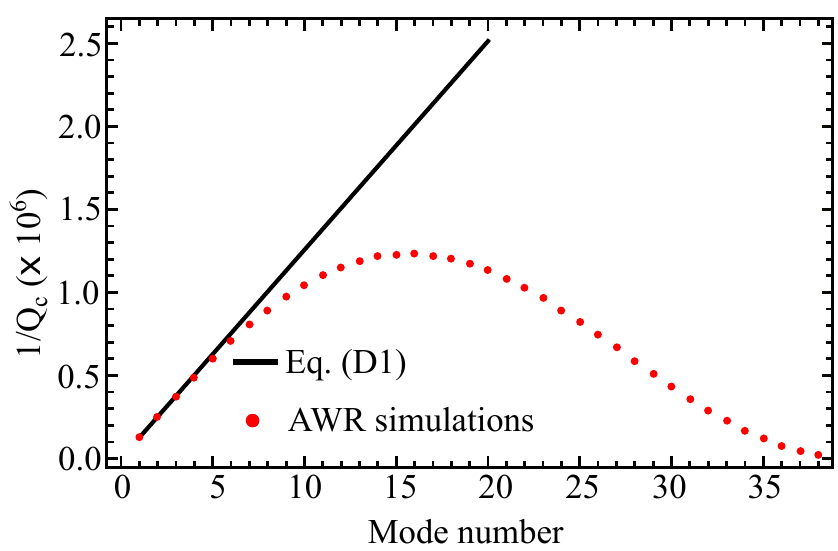}
   \caption{Solid line is $1/Q_c$ expression of a continuous transmission line resonator, as described by Eq.~(\ref{eq:QcTrans}); note that this is plotted as a continuous function for clarity, even though $n$ is limited to integer values. The red points are the $1/Q_c$ values for a 40-cell right-handed discrete transmission line resonator extracted from a circuit simulation using AWR Microwave Office. The parameters for $1/Q_c$ described by Eq.~(\ref{eq:QcTrans}) are: $\omega_0/2\pi=1$~GHz, $Z_0=50\,\Omega$, $C_c=1$~fF. The parameters of the discrete RHTL resonator are: $L_r=0.625$~nH, $C_r=250$~fF, $C_c=1$~fF.
\label{fig:Qc-1VSn}}
\end{figure}

We can use the expressions derived in the previous appendices to investigate the coupling loss for a discrete transmission line resonator. For a resonator formed from a continuous transmission line with characteristic impedance $Z_0$ equal to the load impedance, when 
$\omega_n C_c Z_0 \ll 1$, the coupling loss is given by~\cite{schuster2007circuit}
\begin{equation}
\frac{1}{Q_c}=\frac{4Z_0^2C_c^2{\omega_n}^2}{n\pi}=\frac{4nZ_0^2C_c^2{\omega_0}^2}{\pi}\label{eq:QcTrans},
\end{equation}
where we make use of the fact that such a transmission line will have a linear dispersion, resulting in harmonic resonances $\omega_n = n \omega_0$. In contrast, we use AWR Microwave Office to simulate a discrete transmission line resonator and extracted $Q_c$ for all of the resonance modes. As plotted in Fig.~\ref{fig:Qc-1VSn}, there are dramatic differences in the variation of $1/Q_c$ with $n$ for all but the lowest $n$ modes.


One approach to extract $Q_c$ for each mode of a discrete transmission line resonator is to use Eq.~(\ref{eqn:S21}) for $S_{21}(\omega)$ and fit Lorentzians to each resonance peak to determine the linewidth. Alternatively, it would be useful to derive a closed-form expression for $Q_c$ for more efficient evaluation of circuits. This can be done by considering the equivalent LC resonant circuit for each mode of a discrete transmission line resonator then mapping this onto the expression for $Q_c$ for the simple LC circuit. We do this below -- first for an RHTL resonator, then for a resonator based on an LHTL.

For a basic LC circuit with symmetric input and output coupling capacitors $C_c$ to load and source resistors $R_0$ and resonance frequency $\omega_0^{LC}$, when $\omega_0^{LC} C_c R_0 \ll 1$, then $Q_c$ is given by~\cite{Goppl08,Oconnell08}

\begin{equation}
Q_c=\frac{\tilde{C}}{2\omega_0^{LC} C_c^2R_0}\label{eq:ExternalQ},
\end{equation}
where $\tilde{C}$ is the capacitance of the LC circuit. For a discrete RHTL or LHTL resonator, if we replace $\omega_0^{LC}$ with $\omega_n$ for mode $n$ and determine $\tilde{C}$ corresponding to mode $n$, we can then use this expression to compute $Q_c$.

For an ideal RHTL, we first compute the total electric field energy stored in the RHTL resonator
\begin{equation}
E_{e}^{RHTL}=\frac{1}{2}\sum\limits_{m=-N}^{-1}C_r|v_m|^2\label{eq:EeRHTL1}.
\end{equation}

\begin{figure}[b]
\centering
\includegraphics[width=3.35in]{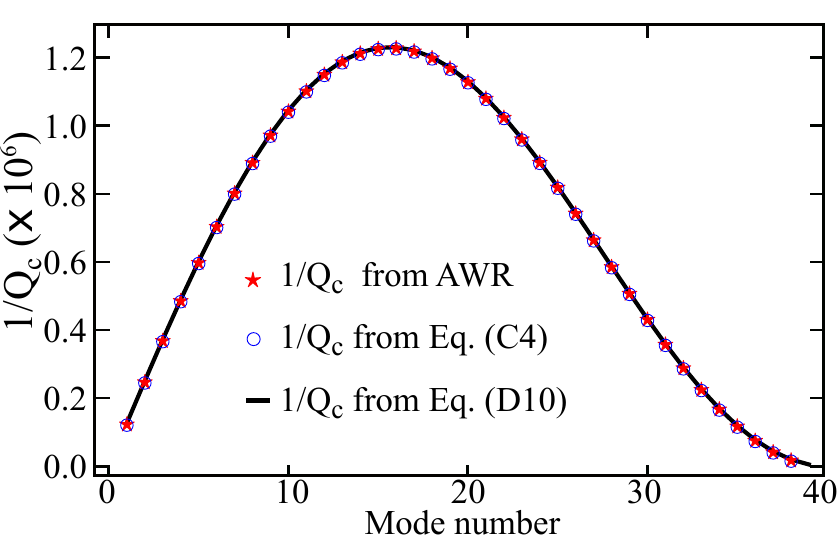}
  \caption{Coupling loss comparison for an RHTL resonator for analytic expression from Eq. (D10), AWR Microwave Office circuit simulation, and linewidth extraction for $S_{21}(\omega)$ expression from Eq.~(\ref{eqn:S21}). 
The parameters for the RHTL are: $C=250$~fF, $L=0.625$~nH, $C_c=1$~fF, $N=40$. 
\label{fig:$RHL Q_C theory$ plot comparison}}
\end{figure}

\noindent When $C_c$ is small, we can take the limit $Z_L=\left(1/i\omega_nC_c+R_0\right)\to\infty$, so that $\Gamma\to e^{-i k \Delta x}$, then using Eqs.~(\ref{eq:VmwithGamma}) and (\ref{eq:A20}), the voltage at each unit cell $m$ is
\begin{equation}
v_m=V_0^+\left[e^{-i\frac{mn\pi}{N}}+e^{i\frac{(m-1)n\pi}{N}}\right]\label{eq:vm}.
\end{equation}
Next, we substitute Eq.~(\ref{eq:vm}) into Eq.~(\ref{eq:EeRHTL1}) and equate the result to the total electric field energy stored in the equivalent LC circuit for a voltage $V_{in}$ across the capacitor
\begin{equation}
\frac{1}{2}\sum_{m=-N}^{-1} C_r |v_m|^2 = N C_r (V_0^+)^2 = \frac{1}{2}\tilde{C}|V_{in}|^2.
\end{equation}
In order to relate $V_0^+$ with $V_{in}$, we recognize that the voltage at the input cell $m=-N$ must be $V_{in}$. Then, using Eqs.~(\ref{eq:A20})-(\ref{eqn:V-N}), we can write
\begin{equation}
|V_{in}|^2 = |V_{-N}|^2 = 4(V_0^+)^2 \cos^2\left( \frac{n\pi}{2N}\right).
\end{equation}
Thus, we extract the appropriate $\tilde{C}$ for mode $n$ of an RHTL resonator, which we label $\tilde{C}_{RHTL}$
\begin{equation}
\tilde{C}_{RHTL}=C_r\frac{N}{2\cos^2\left(\frac{n\pi}{2N}\right)}.
\end{equation}
Combining this result with Eq.~(\ref{eq:QcTrans}) and the RHTL dispersion relation from Eq.~(\ref{eq:RHdispersion}), we obtain
\begin{equation}
Q_{c}^{RHTL}(n)=\frac{C_r^2Z_0N}{8C_c^2R_0\sin\left( \frac{n\pi}{2N}\right)\cos^2\left( \frac{n\pi}{2N}\right)}\label{eq:QCRHL}.
\end{equation}
In Fig.~\ref{fig:$RHL Q_C theory$ plot comparison} we use the analytic expression for $Q_c^{RHTL}$ from Eq.~(\ref{eq:QCRHL}) with the $1/Q_c$ values vs. $n$ for an RHTL resonator from the AWR Microwave Office circuit simulation presented earlier in Fig.~\ref{fig:Qc-1VSn}, as well as linewidth fits extracted from the resonance peaks using Eq.~(\ref{eqn:S21}) for $S_{21}(\omega)$. There is excellent agreement between all three approaches.

For completeness, the inductance of the LC oscillator corresponding to mode $n$ of an RHTL resonator is determined by $1/\sqrt{\tilde{L}\tilde{C}}=\omega_n$, leading to
\begin{equation}
\tilde{L}_{RHTL}=\frac{L_r}{2N\tan^2\left(\frac{n\pi}{2N}\right)}\label{eq:effectiveLRHL}.
\end{equation}

We apply a similar approach to compute the coupling loss for an LHTL resonator, but with Eq.~(\ref{eq:LHdispersion}) for the appropriate dispersion relation, resulting in
\begin{equation}
Q_c^{LHTL}=\frac{Z_0C_l\sin\left(\frac{n\pi}{2N}\right)\tilde{C}_{LHTL}}{C_c^2R_0}\label{eq:QCLHL}.
\end{equation}
In order to compute $\tilde{C}_{LHTL}$, we recognize that all of the capacitors $C_l$ are in series in the LHTL, while the cell voltage $v_m$ is defined relative to ground. Thus, the voltage across the capacitor in cell $m$ is $v_{m-1}$-$v_m$ and the electric field energy stored in the LHTL resonator is
\begin{equation}
\begin{split}
E_e^{LHTL}=&\frac{1}{2}\sum\limits_{m=-N}^{-1}C_l|v_{m-1}-v_m|^2\\
=&4N C_l (V_0^+)^2N\sin^2\left( \frac{n\pi}{2N} \right)\label{eq:EeLHL1},
\end{split}
\end{equation}
using Eq.~(\ref{eq:vm}) for $v_m$.
Thus, the effective capacitance of the equivalent LC circuit for an LHTL resonator is
\begin{equation}
\tilde{C}_{LHTL}=\frac{2N\sin^2 \left( \frac{n\pi}{2N} \right)}{\cos^2\left( \frac{n\pi}{2N} \right)}C_l.
\end{equation}
Substituting this result for $\tilde{C}_{LHTL}$ in Eq.~(\ref{eq:QCLHL}) leads to
\begin{equation}
Q_c^{LHTL}(n)=\frac{2N Z_0 C_l^2 \sin^3 \left( \frac{n\pi}{2N} \right)}{R_0 C_c^2 \cos^2 \left( \frac{n\pi}{2N} \right)\label{eq:QcLeftwithZ0}},
\end{equation}
which, when inverted, reduces to Eq.~(\ref{eq:QcLHL1}) for $Z_0=R_0$.
As with the RHTL resonator analysis, we use Eq.~(\ref{eq:QcLeftwithZ0}) to compare the coupling loss for an LHTL resonator computed with this analytic expression with that obtained from an AWR Microwave Office circuit simulation and linewidth fits extracted from the full expression for $S_{21}(\omega)$ from Eq.~(\ref{eqn:S21}). Again, the agreement between the three approaches is quite good.
\noindent Also for completeness, the inductance of the equivalent LC oscillator for mode $n$ of an LHTL resonator can be found as before to be
\begin{equation}
\tilde{L}_{LHTL}=\frac{2L_l\cos^2\left(\frac{n\pi}{2N}\right)}{N}.
\end{equation}
\begin{figure}[b]
\centering
\includegraphics[width=3.35in]{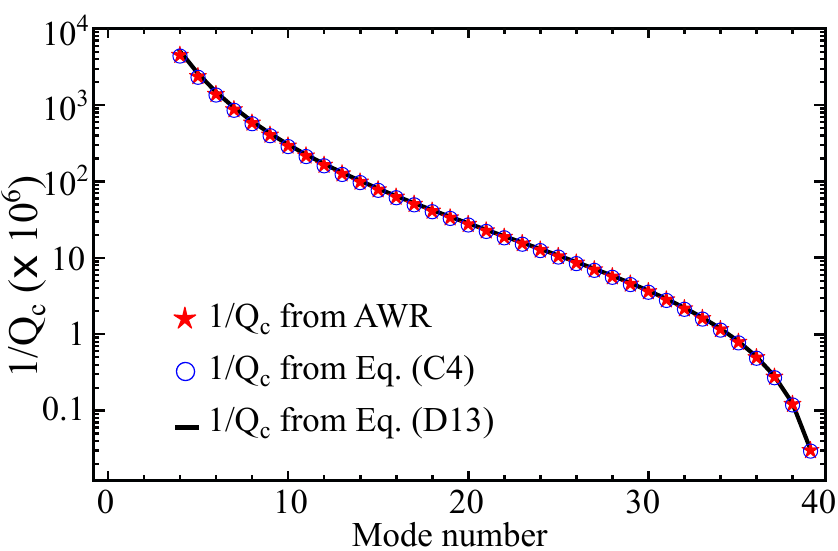}
  \caption{Coupling loss comparison for an LHTL resonator for analytic expression from Eq.~(\ref{eq:QcLeftwithZ0}), AWR Microwave Office circuit simulation, and linewidth extraction for $S_{21}(\omega)$ expression from Eq.~(\ref{eqn:S21}).  The parameters for the LHTL are: $C=250$~fF, $L=0.625$~nH, $C_c=10$~fF, $N=40$. 
\label{fig:LHL Q_C theory plot comparison}}
\end{figure}
\begin{figure}[htb]
\centering
\includegraphics[width=3.35in]{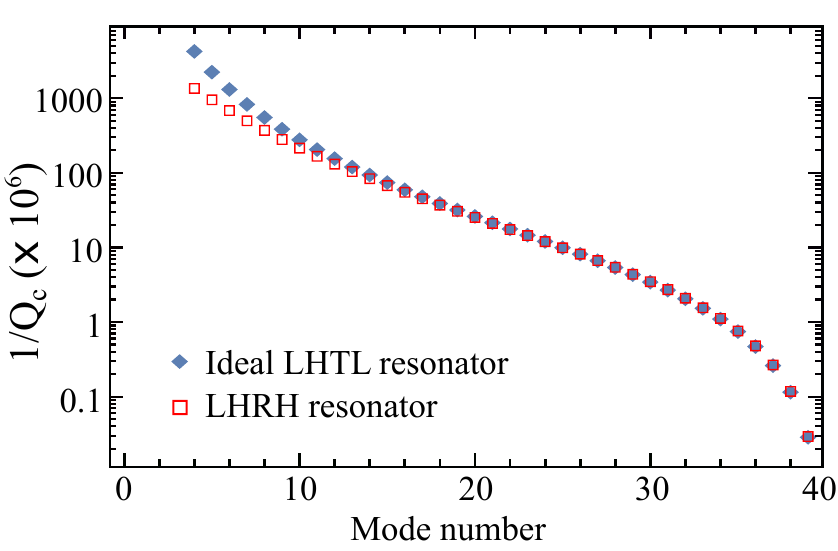}
  \caption{Coupling loss comparison between calculation for an ideal LHTL resonator using Eq.~(\ref{eq:QcLeftwithZ0}) with an LHRH resonator with non-zero stray reactances from linewidth extractions using Eq.~(\ref{eqn:S21}).  The parameters for the LHTL are: $C=250$~fF, $L=0.625$~nH, $C_c=10$~fF, $N=40$ $C_r=16.211389$~fF, $L_r=0.0334947$~nH, corresponding to self resonance frequencies of 50 GHz and 55 GHz. 
\label{fig:hybrid and LHL Q_C plot comparison}}
\end{figure}

Our analysis thus far in this appendix  assumes ideal RHTL or LHTL resonators with no stray reactance. For a realistic circuit that does include such parasitic effects, these stray reactances can indeed be accounted for, but the expressions become rather unwieldy. To test the effects of neglecting the stray reactances in this analysis, in Fig.~\ref{fig:hybrid and LHL Q_C plot comparison} we compare the coupling loss vs. $n$ for an ideal LHTL resonator computed from Eq.~(\ref{eq:QcLeftwithZ0}) with coupling loss values obtained from the complete $S_{21}(\omega)$ expression from Eq.~(\ref{eqn:S21}) using realistic values of stray reactance $L_r$ and $C_r$ included. As the comparison shows, the analytic expression from Eq.~(\ref{eq:QcLeftwithZ0}) for an ideal circuit agrees reasonably well with the realistic LHRH resonator for mode numbers beyond $\sim 10$. Only the lowest $n$ (highest frequency) modes have any significant deviation. Thus, the simple analytic expression from Eq.~(\ref{eq:QcLeftwithZ0}) can be used for estimating coupling losses for LHTL resonators for all but the highest frequency modes, where a numerical extraction of linewidths using Eq.~(\ref{eqn:S21}) should be used instead.

\section{Spatial variations in charge and current distributions in LHTL resonators}
\label{appendix:phase}
\begin{figure}
\center
\includegraphics[width=3.35in]{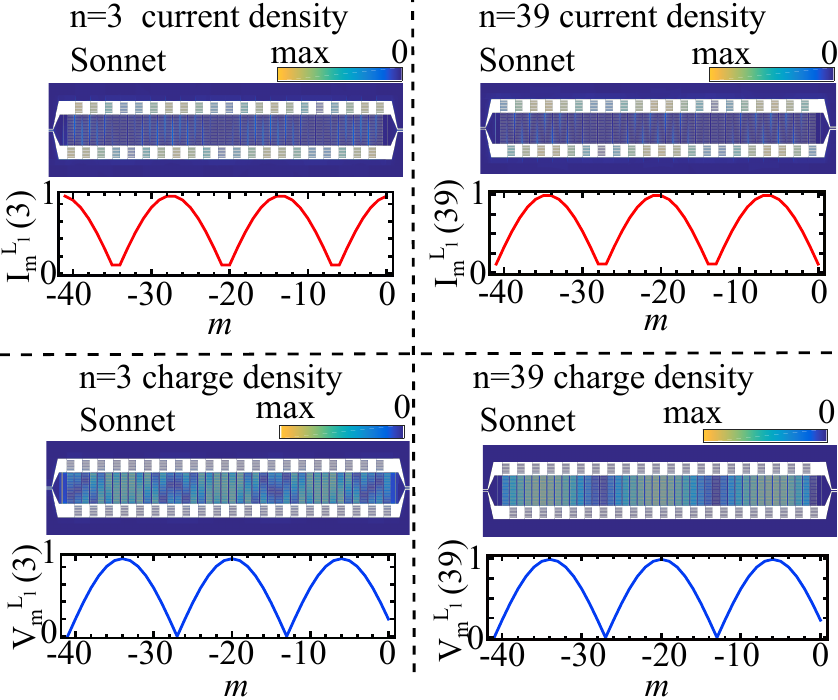}
\caption{Standing wave patterns for modes $n=3$ and 39 of 42-cell LHTL resonator from Sonnet simulations of current density and charge density distributions compared with expressions for current in inductors (voltage across capacitors) from Eq.~(\ref{eq:I_Llm}) [Eq.~(\ref{eq:V_Clm})] normalized by the maximum value for each mode of current $i_{0}(n)$ (voltage $v_{0}(n)$) so $I_m^{L_l}(n)=|i_m^{L_l}(n)/i_0(n)|$ ($V_m^{C_l}(n)=|v_m^{C_l}(n)/v_0(n)|$).
\label{fig:volcurPattern}}
\end{figure}

As discussed in Sec.~\ref{sec:LSM imaging of mode structure}, when we image our metamaterial resonator while exciting the various modes, we observe both the microwave current and charge density distributions and these are consistent with Sonnet simulations (Fig.~\ref{fig:ZoomedIn}). In contrast to standing waves on a continuous transmission line resonator, where the antinodes in the current will always match the nodes in the voltage and vice versa, for our images of the $n=40$ mode for our metamaterial resonator, we observe the current density and charge density distributions to be spatially in phase. In this Appendix, we show that this behavior is expected for a discrete transmission line resonator when exciting a high-$n$ mode; while for a low-$n$ mode, the two distributions are spatially out of phase, as with a conventional continuous resonator. We begin by revisiting the treatment in Appendix~\ref{sec:Impedance of general terminated discrete transmission line}, where Eqs.~(\ref{eq:VmwithGamma}) and (\ref{eq:ImwithGamma}) give the voltage and current at cell $m$. Here, we write these expressions explicitly for mode $n$
\begin{align}
v_{m}(n)&=V_0^+\left[e^{-im\frac{n}{N}\pi}+e^{i(m-1)\frac{n}{N}\pi}\right]\label{eq:E1}\\
i_{m}(n)&=I_0^+\left[e^{-i(m+\frac{1}{2})\frac{n}{N}\pi}-e^{i(m-\frac{1}{2})\frac{n}{N}\pi}\right]\label{eq:E2}.
\end{align}
As discussed in Appendix~\ref{sec:couplingloss}, the voltage across each capacitor of the LHTL is: $v_{m}^{C_l}=v_{m-1}-v_{m}$ and the current flowing through each inductor is $i_{m}^{L_l}=i_{m-1}-i_{m}$. Thus, on resonance for mode $n$, the voltage across the capacitor and current through the inductor for cell $m$ are
\begin{align}
\begin{split}
v_{m}^{C_l}(n)&=2V_0^+\sin\left[\left(m-1\right)\frac{n\pi}{N}\right]\\
&\times\left[\sin\left(\frac{n\pi}{N}\right)+i-i\cos\left(\frac{n\pi}{N}\right)\right]
\end{split}\label{eq:V_Clm}\\
\begin{split}
i_{m}^{L_l}(n)&=2I_0^+\cos\left[\left(m-\frac{1}{2}\right)\frac{n\pi}{N}\right]\\
&\times\left[1-\cos\left(\frac{n\pi}{N}\right)+i\sin\left(\frac{n\pi}{N}\right)\right].
\end{split}\label{eq:I_Llm}
\end{align}
In order to compare with our Sonnet simulations of the charge density, which is related to the voltage across the capacitors, and the current density, dominated by the current through the inductors, we take the magnitude of Eqs.~(\ref{eq:V_Clm})-(\ref{eq:I_Llm}). If we start by examining the low-$n$ limit, for mode $n=1$, we obtain
\begin{align}
\left|v_{m}^{C_l}(1)\right|&\approx 2 V_0^+\left|\left(\frac{\pi}{N}\right)\sin\left[\frac{\left(m-1\right)\pi}{N}\right]\right|\label{eq:Vclm1}\\
\left|i_{m}^{L_l}(1)\right|&\approx 2 I_0^+\left|\left(\frac{\pi}{N}\right)\cos\left[\frac{\left(m-\frac{1}{2}\right)\pi}{N}\right]\right|\label{eq:Iclm1}.
\end{align}
From Eqs.~(\ref{eq:Vclm1}) and (\ref{eq:Iclm1}), it is clear that the voltage and current standing wave patterns are spatially out of phase, as one would expect for a continuous transmission line resonator. However, if we examine the large-$n$ limit, for mode $n=N-1$, we obtain
\begin{align}
\left|v_{m}^{C_l}(N-1)\right|&\approx 4 V_0^+\left|\sin\left[\left(\frac{N-1}{N}\right)\left(m-1\right)\pi\right]\right|\label{eq:E5}\\
\left|i_{m}^{L_l}(N-1)\right|&\approx 4 I_0^+\left|\sin\left[\left(\frac{N-1}{N}\right)\left(m-1\right)\pi\right]\right|\label{eq:E6},
\end{align}
thus, the voltage and current standing wave patterns are now in phase spatially. This behavior is also reflected in our Sonnet simulations. In Fig.~\ref{fig:volcurPattern}, we compare the standing wave patterns for mode $n=3$ and 39, computed from the magnitude of Eqs.~(\ref{eq:V_Clm})-(\ref{eq:I_Llm}) with the corresponding Sonnet simulations of the charge density and current density.

\begin{figure}[htb]
\centering
\includegraphics[width=3.35in]{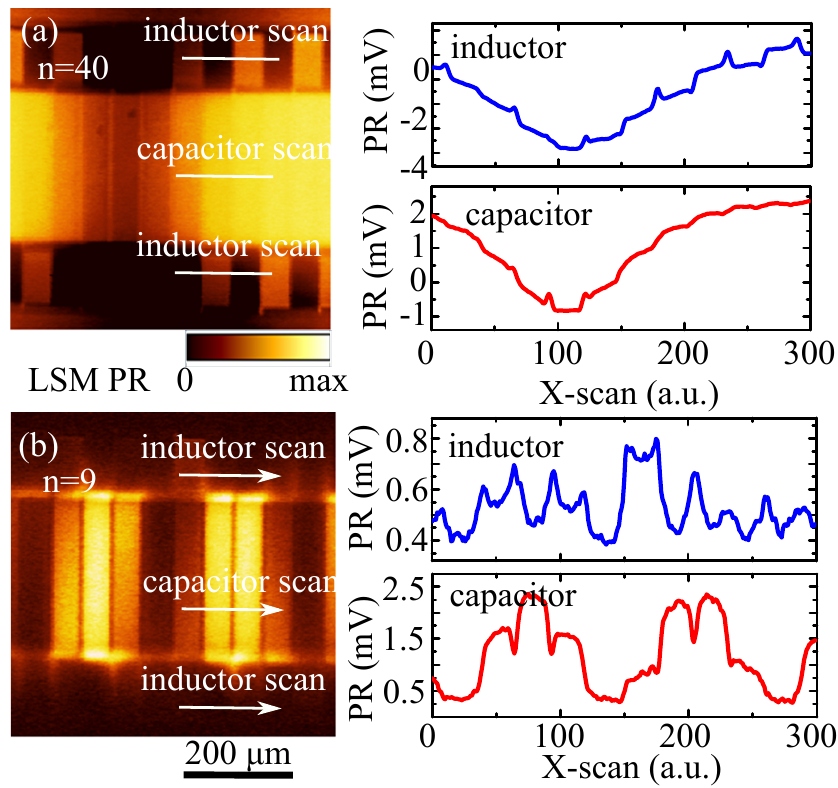}
  \caption{High-resolution LSM images of 11 cells near center of metamaterial resonator for (a) $n=40$ and (b) $n=9$ modes. Bright (dark) regions correspond to large (small) PR signal. Plots to the right of image correspond to linecuts as indicated on the images. For the inductor linecuts, the traces through the top and bottom inductors are added together.
\label{fig:LSMPhasePattern}}
\end{figure}

In addition to the Sonnet simulations and Eqs.~(\ref{eq:E5})-(\ref{eq:E6}), we can also observe this behavior in the relative positions of the standing-wave antinodes in the LSM images, as mentioned earlier for Fig.~\ref{fig:ZoomedIn}. In Fig~\ref{fig:LSMPhasePattern}, we present high-resolution LSM images using a smaller laser spot size (5~$\mu$m) of the middle 11 cells of the metamaterial resonator for modes $n=40$ and 9. For each image, we include plots of linecuts across the inductors at the top and bottom as well as the center of the capacitors. For $n=40$ (3.44~GHz), clearly the node in the LSM signal is in the same position for the inductor and capacitor; similarly, the antinodes line up as well. For $n=9$ (15.82~GHz), although the PR signal in the inductors is weaker for this high-frequency excitation, the antinodes and nodes no longer line up between the capacitor and inductor linescans.

We  follow a similar analysis for a discrete RHTL resonator. Again, starting from Eqs.~(\ref{eq:E1})-(\ref{eq:E2}), we derive similar expressions for mode $n$ for the current through the inductor $m$, $i_m^{L_r}$(n), and voltage across the capacitor $m$, $v_m^{C_r}(n)$. For the $n=1$ mode, we obtain
\begin{align}
\left|v_{m}^{C_r}(1)\right|&\approx 2V_0^+\left|\cos\left(\frac{m\pi}{N}\right)\right|\\
\left|i_{m}^{L_r}(1)\right|&\approx 2I_0^+\left|\sin\left(\frac{m\pi}{N}\right)\right|.
\end{align}
Thus, the voltage antinodes will line up with the current nodes and vice versa, as one would observe with a continuous transmission line resonator. Next, we can examine the opposite limit, where for $n=N-1$, we have
\begin{align}
\left|v_{m}^{C_r}(N-1)\right|&\approx 2 V_0^+\left|\sin\left(\frac{m\pi}{N}\right)\right|\\
\left|i_{m}^{L_r}(N-1)\right|&\approx 2 I_0^+\left|\sin\left(\frac{m\pi}{N}\right)\right|.
\end{align}
As with the LHTL resonator for high $n$, the voltage and current standing-wave patterns are spatially in phase. Of course, for the RHTL resonator, this occurs for the highest frequency modes, while for the LHTL resonator, it is the lowest frequency modes.
\section{Effects of imperfect grounding}
\label{appendix:ImpGround}
In our experimental setup for measuring $S_{21}(f)$, the Nb ground plane of the chip is connected to the copper ground plane of our sample holder through a series of short aluminum wire bonds around the perimeter of the chip. In addition, there are multiple jumper wirebonds across the LHTL and the CPW segments on either end to ensure that the different sections of ground plane are as close to a uniform equipotential as possible. However, as is well known with superconducting thin-film microwave circuits~\cite{Wenner2011}, the nonzero self-inductance of the wirebonds can lead to imperfect grounding over the entire area of the device. Here, we attempt to simulate the effects of such imperfect grounding conditions using Sonnet to study how this affects the spectrum and coupling quality factor of the metamaterial resonances.

The typical grounding setup in Sonnet involves defining a grounding box and attaching the metal layer boundaries of the device to the sides of the grounding box. The grounding box and the device are normally both rectangular to fit together exactly so that the simulated metal layer is perfectly grounded. In addition to the ideal grounding scenario described above, which we refer to as ``grounded", we consider three other levels of ground connection quality where we extend the ground box so that it has a $200-\mu{\rm m}$ gap around the perimeter of the chip. We then add different numbers (12, 3, and 0) of $6 \times 200\,\mu{\rm m}$ metal strips to join the edges of the chip to the ground box. We note that these simulated connections are smaller than our actual wirebonds, which have a diameter of $32\,\mu{\rm m}$ and a typical length of $\sim 500\,\mu{\rm m}$ in order to enhance the effects of reduced ground connection quality.

\begin{figure}[b]
\center
\includegraphics[width=3.35in]{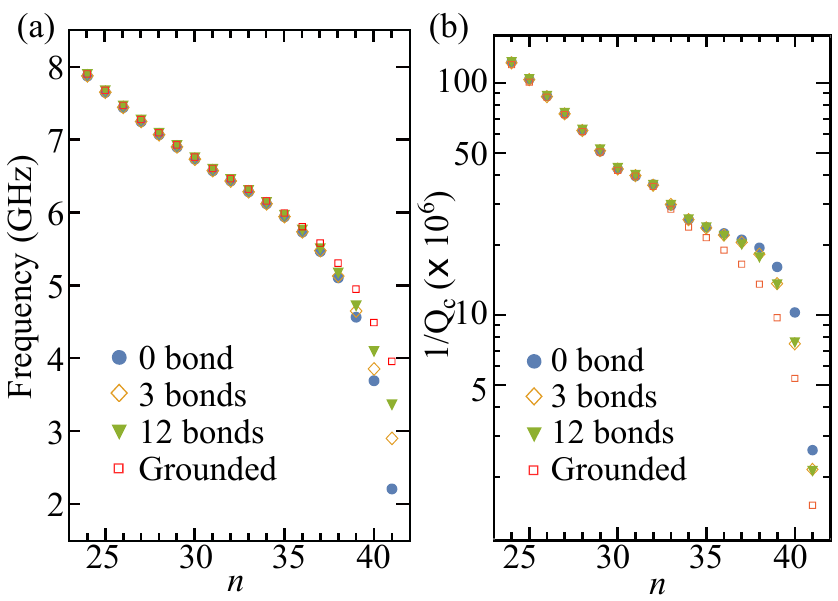}
  \caption{Sonnet simulations for different grounding configurations for a 42-cell LHTL resonator for (a)~resonance frequency vs. $n$, (b)~coupling loss $1/Q_c$ vs. $n$.
\label{Fig:ImperfectGround}}
\end{figure}

In Fig.~\ref{Fig:ImperfectGround}, we plot the results of these Sonnet simulations of a 42-cell LHTL resonator for the various grounding configurations where we compare the frequency of the resulting modes as well as the coupling loss for each mode from $n=24-41$. We observe that the mode frequencies are only influenced significantly for the highest $\sim 5$ mode numbers (and hence lowest frequencies); while the coupling loss of only the highest $\sim 10$ modes is impacted by the ground quality. We do not plot modes $n < 24$ (that is, higher frequency modes) here since there are no discernible differences for the different grounding configurations. Generally, for the highest $n$ and lowest frequency modes, the worse the grounding quality, the lower the frequency of the mode and the higher the coupling loss. These simulations provide a qualitative explanation for most of the differences in our measured mode frequencies and coupling losses and the Sonnet simulations with ideal grounding and the calculated theoretical values based on the circuit parameters, as presented earlier. We note that in the case of the microwave transmission measured during the LSM imaging, the downward frequency shifts of the highest $n$ and lowest frequency modes are even larger than in our ADR measurements. In order to allow the LSM laser spot to access all of the portions of the metamaterial, no jumper wirebonds are used across the metamaterial or CPW segments, which likely degraded the ground integrity. The reduced mode frequencies are consistent with the Sonnet simulations performed here.

\section{Effects of staggered inductor layout}
\label{sec:staggered}
As described in Sec.~\ref{sec:Fab}, we chose to use a staggered layout for the inductors in our metamaterial resonators, as shown in the device images of Fig.~\ref{fig:fab}. In this Appendix, we use Sonnet simulations to examine the effects of the inductor layout configuration on the resonator spectrum. We compare the simulated spectrum for our measured device with that for a similar metamaterial resonator, but with a non-staggered layout with all of the inductors on one side of the metamaterial. For this non-staggered layout, we maintained the same capacitor parameters for the finger width, length, spacing, and number of pairs; the meander-line inductor linewidth and number of turns are also matched to our original layout. However, in order to avoid adjacent inductors from overlapping, the interdigitated capacitor spines are widened so that the unit cell size increased by $6\,\mu{\rm m}$. Thus, the total length of the non-staggered metamaterial resonator must increase by 252 $\mu$m beyond that of our measured device.
\begin{figure}[t]
\centering
\includegraphics[width=3.35in]{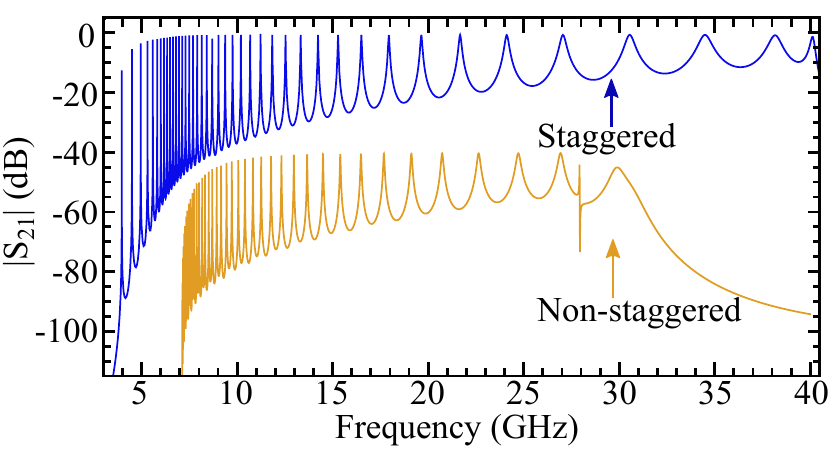}
  \caption{Sonnet simulations of $S_{21}(f)$ for non-staggered (orange, with -40 dB offset for clarity) and staggered (blue) inductor layouts. Note that the blue curve is the same as in Fig.~\ref{fig:sonnet}. 
\label{fig:staggerspectrum}}
\end{figure}


In Fig.~\ref{fig:staggerspectrum} we compare the simulated $S_{21}(f)$ spectra for the original staggered and non-staggered inductor configurations. Although both spectra follow the overall pattern expected for an LHRH resonator, as presented earlier, there are clear differences between the spectra for the two inductor layouts. In the case of the non-staggered inductor configuration, the infrared cut-off frequency is about 2~GHz higher than that for the staggered inductor layout used in our experiments. In addition, the modes just above $\omega_{IR}/2\pi$ have the highest density, as expected from the theoretical treatment of the LHTL and LHRH resonator spectra from Sec. II. This is in contrast to the staggered inductor spectrum, where the modes reach their highest density for a frequency about 2 GHz above $\omega_{IR}/2\pi$. This discrepancy between the spectra for the two inductor configurations can also be seen in Fig.~\ref{fig:Freq-mode-all}, where the mode frequencies of the non-staggered inductor configuration follow the theoretical dispersion relation closely, while the lowest frequency and highest $n$ modes of the staggered inductor configuration fall below the theoretical curve. Based on the Sonnet simulations, we conclude that the distribution of the currents through the staggered inductors for the shortest wavelength modes cause deviations from ideal lumped-element behavior, leading to reduced resonance frequencies.

Besides the differences near the infrared cutoff, the spectra for the two inductor configurations also deviate at high frequencies. The gap beyond mode $n=1$ occurs about 12~GHz lower in frequency for the non-staggered inductor configuration compared to the staggered layout. Thus, although the simulations of the non-staggered configuration lead to closer agreement with the theoretical dispersion relation for medium- and high-$n$ modes, the arrangement with all of the inductors on one side of the metamaterial leads to excess stray reactances $L_r$ and $C_r$, resulting in the decreased gap at high frequency.

In addition to differences in the mode frequencies, we can also use the Sonnet simulations to study the effect of the inductor configuration on the coupling loss. As shown in Fig. 15 where we plot $1/Q_c$ vs. $n$ for the experimental measurements, AWR circuit simulations of an ideal LHTL resonator, and Sonnet simulations for both inductor configurations, the Sonnet simulation for non-staggered inductors is reasonably close to the theoretical dependence derived in Appendix D. In contrast, the staggered inductor simulation from Sonnet shows excess coupling loss for modes beyond $n \sim 30$, consistent with our experimentally measured values. Again, this is consistent with deviations from ideal lumped-element behavior for the short wavelength, high-$n$ modes, which impact the coupling loss dynamics as well as the mode frequencies.

\bibliographystyle{apsrev4-1-newv2}
\bibliography{meta-mode-Paper-Refv1}
\end{document}